\documentclass{JHEP3}
\usepackage{latexsym}
\usepackage{graphics}
\usepackage{epsfig}
\usepackage{multirow}
\usepackage{amssymb,amsmath}

\newcommand{\ie}{{i.e.}~}
\newcommand{\eg}{{e.g.}~}
\newcommand{\zbbd}{\ensuremath{gg \to Z^0/\gamma^* b \bar{b} \to f \bar
f b \bar{b} ~}}
\newcommand{\zdec}{\ensuremath{Z^0/\gamma^* \to f \bar f ~}}
\newcommand{\zz}{\ensuremath{Z^0/\gamma^*\,}}
\newcommand{\zbb}{\ensuremath{gg \to Z^0/\gamma^* b \bar{b} ~}}
\newcommand{\zbbp}{\ensuremath{AB \to Z^0/\gamma^* b \bar{b} ~}}

\title{A Consistent Prescription for Combining Perturbative Calculations and
Parton Showers in Case of Associated $\mathbf{Z^0 b \bar{b}}$ Hadroproduction  }

\author{Borut Paul Kersevan$^{a,b}$ \\ $^a$ Jozef Stefan Institute,
  Jamova cesta 39, SI-1000 Ljubljana, Slovenia \\
$^b$  Faculty of Mathematics and Physics,University of Ljubljana,\\ Jadranska 19a, SI-1000 Ljubljana, Slovenia\\
E-mail: \email{borut.kersevan@ijs.si}}
\author{Ian Hinchliffe\thanks{Work supported  by the Director, Office of Science, Office of
High Energy Physics, of the U.S.\ Department of Energy under Contract
DE-AC02-05CH11231.}\\ Lawrence Berkeley National Laboratory, Berkeley, CA , 94720 USA.\\
E-mail: \email{I\_Hinchliffe@lbl.gov} }
\author{Liza Mijovic \\
 Faculty of Mathematics and Physics,University of Ljubljana,\\ Jadranska 19a, SI-1000 Ljubljana, Slovenia\\
E-mail: \email{liza.mijovic@ijs.si}}

\abstract{This paper presents the method of combining parton shower
formalism with perturbative calculations (matrix elements) in form of
a Monte-Carlo algorithm for the process \zbb, consistently including the heavy quark masses and
overlap removal. }


\preprint{arXiv:0803.4132 [hep-ph] }

\begin{document}

\section{Introduction}

The \zbbd process is of high experimental interest in view of the forthcoming
LHC experiments, since it \eg represents an irreducible background to
the `gold-plated' Higgs  channel $H \to ZZ^{(*)} \to 4  ~{\rm leptons}$
(see \eg \cite{ATL-PHYS-TDR}). Historically, the process has been fully
calculated at tree level by Kleiss and Stirling \cite{Kleiss1990}
and for the first time successfully implemented inside the AcerMC
\cite{Kersevan:2002dd} Monte-Carlo generator. In the present time there is a
plethora of other Monte-Carlo generators implementing this process in
various advanced fashions (see \eg \cite{Schalicke:2005nv}), however a few
issues still need to be resolved in a consistent fashion:
\begin{itemize}
\item At tree level the \zbbd process is actually  NNLO
in $\alpha_s$ with respect to the order $\alpha_s^0 $ `pure' Drell-Yan process $b \bar{b}
\to \zdec$. The same final state can thus be achieved by using the
Sudakov parton showering procedure, which by definition re-sums the
large logs of the order $\alpha_s \ln{M^2_Z/m^2_b}$ which burden the higher
order corrections as the \zbb and might thus be better at least in the
'intermediate' kinematic regions (see \eg \cite{Sjostrand:2006za} where the
\zbbd process was actually removed with this argument). While the
argument by all means stands it would nevertheless be preferable to
have the processes consistently combined in orders $\alpha^n_s,\,
n=0,1,2$, achieving the undisputed validity over the whole
phase-space.   
\item The b-quarks are reasonably massive; still it is customary to
treat all partons incoming to the hard process as massless, which is
strictly speaking consistent only if also the final state b-quarks are
also treated as massless. In `full' Monte-Carlo procedures, where the
produced b-quarks are further hadronized into jets, the mass of the
b-quark needs to be present and is thus added in and \emph{ad-hoc}
fashion. Furthermore, neglecting the heavy quark masses of the
incoming quarks has been shown to have an observable impact and
can in fact be consistently added into the Factorization Theorem
\cite{Aivazis:1993kh,Aivazis:1993pi,Olness:1997yc}.
\end{itemize}

The existing prescriptions deal either with massive particles
\cite{Aivazis:1993kh,Aivazis:1993pi,Olness:1997yc} on the level of integrated cross-sections or
with explicit Monte-Carlo algorithms involving light ($\sim$~massless) particles (\eg
\cite{Schalicke:2005nv,Sjostrand:2006za,Bengtsson:1987rw,Chen:2001ci,Catani:2001cc,Mrenna:2003if}
) while a first attempt at the combination of the two was attempted in
the paper \cite{Kersevan:2006fq}, where an algorithm combining the two features
was developed but implemented only in terms of order $\alpha_s^1$
correction while for the \zbb process the order $\alpha_s^2$
combination procedure is needed. The aim of this paper is to show that
the procedure developed in \cite{Kersevan:2006fq} is in fact iterative and can
thus provide a consistent procedure applicable (at least) at
tree-level which can be and is implemented in terms of a full \zbb
Monte-Carlo procedure.

\section{Combining the Perturbative QCD and Sudakov Showering in Massive Hadroproduction} 

\subsection{Theoretical Basis}

\FIGURE{
  \epsfig{file=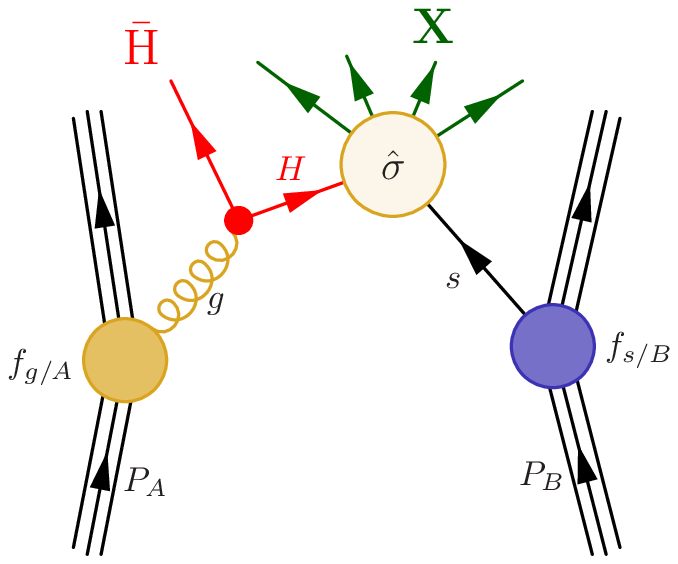,width=4.5cm}
\caption{ Schematic representation of the example of a gluon splitting
$g \to H \bar{H}$ combined with a hard subprocess $\hat{\sigma}$ \label{f:dglap}.}
}

Let us start by repeating the considerations of the Factorization
Theorem
\cite{Collins:1998rz,Altarelli:1977zs,Collins:1977iv,Collins:1981uw,Olness:1989ke},
which states that the hadronic cross cross section $\sigma_{\rm AB \to
  X}$ is related to the perturbatively calculated (\eg by using
Feynman diagram technique) parton level-cross section $\sigma_{\rm ab
  \to X}$ by:
\begin{equation}
\sigma_{\rm AB \to X} = \sum_{a,b} f_{a/A} \otimes \hat{\sigma}_{ab \to X} \otimes
f_{b/B},
\label{e:factheor}
\end{equation}
with $f_{i/I}=f_{i/I}(x,\mu_F)$ denoting the  parton density functions (PDF),
giving a probability that a fully evolved parton $i$ is produced by the
parent hadron $I$, at the factorization scale $\mu_F$ with a certain
energy fraction $x$. The PDFs are convoluted with the \emph{hard}
parton-level cross section $\hat{\sigma}_{ab \to X}$ which can in
general differ from the perturbative (pQCD) parton level-cross section
$\sigma_{\rm ab \to X}$.

Note that the term \emph{hard cross section} in the above application
of the Factorization Theorem demands that the hard cross section
expression of Eq. \ref{e:factheor} is indeed describing the
\emph{hard} ('short-distance', high energy) process only, \ie all the
'long-distance' contributions in form of collinear/mass singularities
and corresponding large logarithms in form of $\alpha_s
\log(\mu_F^2/m^2)$ need have been explicitly subtracted since they are
already included (resummed) in the PDFs
\cite{Collins:1998rz,Aivazis:1993kh,Aivazis:1993pi}. The factorization
scale $\mu_F$ sets the dividing limit between the two kinematic
regimes.

Let us illustrate how this applies to the case studied in this paper:
In a perturbative calculation of order $\alpha_s^{n}$ an incoming
gluon splits to a heavy quark pair $g \to H \bar{H}$; the other
incoming parton we mark as the spectator $s$ (c.f. Figure
\ref{f:dglap}).  The Factorization theorem then gives:
\begin{equation}
\sigma^{(n)}_{\rm AB \to X \bar{H}} = f_{g/A} \otimes \hat{\sigma}^{(n)}_{\rm
 g s \to X \bar{H} } \otimes f_{s/B}\,. \label{e:factheavy}
\end{equation}
Possible summations
and permutations of incoming flavors are omitted and only the
convolution with the parton density functions remains explicit. 

Stipulating that the
(hard/soft) scale $\mu$ is set by the heavy quark kinematics (\eg
heavy quark propagator; other choices are possible) then :
\begin{itemize}
\item If the scale is hard enough, $\mu > \mu_F$ , the pQCD calculation
  need not be modified, \ie 
\begin{equation}
\hat{\sigma}^{(n)}_{\rm g s \to X \bar{H}} =
  \sigma^{(n)}_{\rm g s \to X \bar{H}}~~~~~~~~~(\mu > \mu_F)
\label{e:hlim}
\end{equation}
 in this kinematic region. 
\item if the scale is soft, $ \mu < \mu_F$, the heavy quark production
  in gluon splitting is 'long-distance' and thus already included in
  the PDF $f_{H/I}$. In other words, the incoming gluon cannot be
  resolved at this scale $\mu_F$ and the heavy quark $H$ should be
  considered as the incoming (fully evolved) parton. One
  should thus use an alternative calculation $\hat{\sigma}^{(n-1)}_{H s \to X}$
  with an incoming quark $H$ instead: 
\begin{equation}
\sigma^{(n-1)}_{\rm AB \to X } = f_{H/A} \otimes \hat{\sigma}^{(n-1)}_{\rm
  H s \to X } \otimes f_{s/B}\,~~~~~~(\mu < \mu_F)
\label{e:olness0}
\end{equation}
and correct the perturbative 
calculation correspondingly.  
\end{itemize}

The latter case ($ \mu < \mu_F$) however needs to be explained a bit
further: One should \emph{not} simply set $\hat{\sigma}^{(n)}_{\rm g s \to X \bar{H}}=0$ since 
only the collinear kinematic limit (explicit large logs $\alpha_s
\log(\mu_F^2/m^2)$) is resummed in the $f_{H/I}$, whereas such logs
are only a part (limit) of the full pQCD calculation $\hat{\sigma}^{(n)}_{g s \to X
  \bar{H}}$ and it is only these contributions that need to be removed
in order to prevent double counting of the inclusive $\hat{\sigma}^{(n-1)}_{H s \to X}$
contribution in this kinematic region. Instead one should put:
\begin{equation}
\hat{\sigma}^{(n)}_{\rm g s \to X \bar{H} } = \sigma^{(n)}_{\rm
 g s \to X \bar{H} } - \sigma^{(n)}_{\rm subt},
\end{equation}
which can then be used in the hadronic
cross-section expression:
\begin{equation}
\sigma^{(n)}_{\rm AB \to X \bar{H}} = f_{g/A} \otimes \hat{\sigma}^{(n)}_{\rm
 g s \to X \bar{H} } \otimes f_{s/B}\, = 
f_{g/A} \otimes \sigma^{(n)}_{\rm  g s \to X \bar{H} } \otimes f_{s/B}\, 
- f_{g/A} \otimes \sigma^{(n)}_{\rm subt} \otimes f_{s/B}\, .  
\label{e:olness1}
\end{equation}
One can immediately deduce one property of the subtraction term from Equation \ref{e:hlim}:
\begin{equation}
\sigma^{(n)}_{\rm subt}=0 ~~~~~~~~~(\mu > \mu_F).
\end{equation}

\FIGURE{
     \epsfig{file=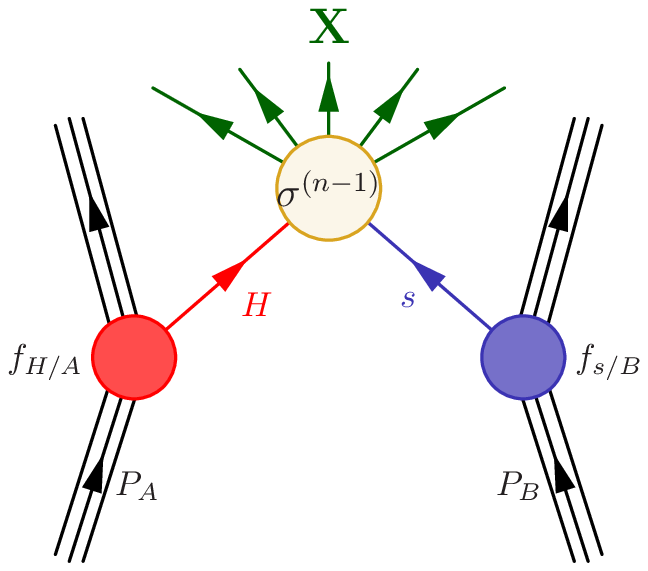,width=4.5cm}\parbox{0.3cm}{\vskip -3.5cm \Large$\oplus$}
     \epsfig{file=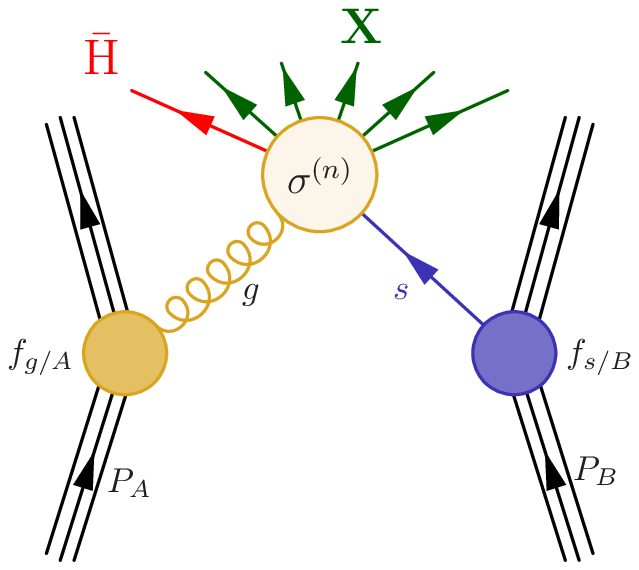,width=4.5cm}\parbox{0.3cm}{\vskip -3.5cm \Large$\ominus$}
     \epsfig{file=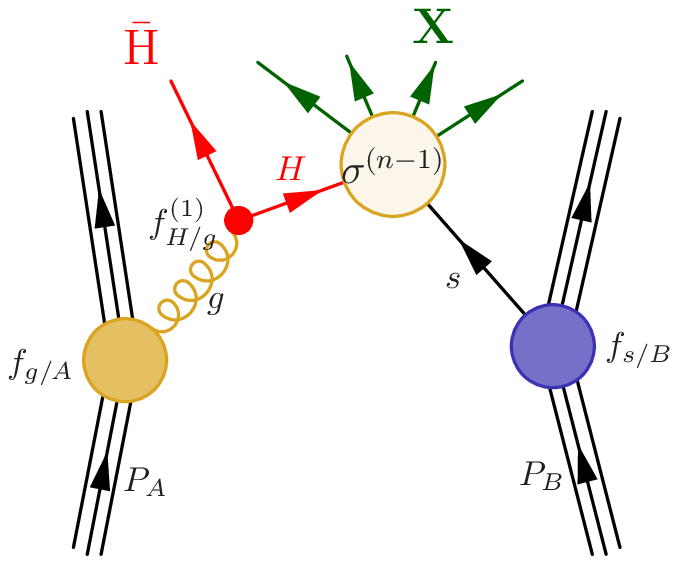,width=4.5cm}
\caption{ Schematic representation of heavy quark H entering the
perturbative calculation at order (n-1) as fully evolved (left) or
participating internally via the splitting of the incoming gluon
at order (n) (middle) with the appropriate subtraction term
(right). The subscript $s$ denotes the other incoming parton (which can also be a
heavy quark) and X the inclusive final state.\label{f:olness}}
}

The full (correct) cross section involving the heavy flavour
excitation in the initial state (category HE in the paper
\cite{Olness:1997yc}) thus needs the addition of the
$\hat{\sigma}^{(n-1)}_{H s \to X}$ hard process to correctly cover the
soft region:
\begin{eqnarray}
\sigma_{\rm AB \to X (\bar{H})}  &=&  
f_{g/A} \otimes \sigma^{(n)}_{\rm  g s \to X \bar{H} } \otimes
f_{s/B}\, \label{e:olness} \\
&-& f_{g/A} \otimes \sigma^{(n)}_{\rm subt } \otimes f_{s/B}\,
\notag \\  
&+& f_{H/A} \otimes \hat{\sigma}^{(n-1)}_{\rm H s \to X} \otimes f_{s/B}\,,~~~~~~(\mu < \mu_F) \notag  
\end{eqnarray}
as illustrated also in Figure \ref{f:olness}.

At this point it should be emphasised that all the  above hadronic
cross-sections are \emph{effectively of the same order} if one
considers that the heavy quark density $f_{H/I}$ is of the effective
order $\alpha_s$ higher with respect to the dominant (gluon or valence
quark) parton distribution functions (see \eg \cite{Olness:1997yc} for
details). Also note that the final state heavy quark $\bar{H}$ need not be
resolved in the final state since both final states $X\bar{H}$ and $X$
from the two hard contributions participate in the full expression of
Equation \ref{e:olness} and the $\bar{H}$ can in the second case
appear only in the soft (parton showering) processes. 

An excellent basis for deriving the desired rules and
expressions is the paper of Olness, Scalise and Tung \cite{Olness:1997yc},
which presents a comprehensive review of the formalism needed for
obtaining the fully infra-red safe hard cross-sections with the
emphasis on isolating and subtracting the divergences related to the
heavy quark terms. The derivation of \cite{Olness:1997yc} (or equivalently \cite{Olness:1989ke}
and our previous paper \cite{Kersevan:2006fq}) gives the expression for the
subtraction term of the order (n) as (please consult the Appendix
\ref{s:app} for details):
\begin{equation}
\sigma^{(n)}_{\rm subt} =  f^{(1)}_{H/g} \otimes \hat{\sigma}^{(n-1)}_{\rm
H s \to X }\, , ~~~~~~(\mu < \mu_F)  
\label{e:olnessubt}
\end{equation}
where the first-order in $\alpha_s$ term $f^{(1)}_{H/g}$ is the  perturbatively calculated
parton distribution function $f^{(1)}_{i/j}$ of a parton \emph{inside another parton}
which is explicitly given as (c.f. \cite{Olness:1997yc,Kersevan:2006fq}) :
\begin{equation} 
f^{(1)}_{i/j}(\xi,\mu) = \frac{\alpha_s(\mu)\,}{2\pi} 
P_{j \to i}(\xi) \ln\left(\frac{\mu^2}{m^2_H}\right) 
\end{equation}  
with $P_{j \to i}$ being the well-known splitting kernels and also 
contains the explicit massive divergence logarithm. 

This is not necessarily the final step one needs to make to be able to
evaluate the hadronic cross sections, since \eg the spectator $s$ can also 
be a gluon splitting to heavy quarks (as is indeed the case in the
studied \zbb process) and thus the whole procedure and calculation of the
subtraction terms needs to be recursively (iteratively) repeated for both n$^{\rm th}$ and (n-1)$^{\rm th}$ order terms in
order to obtain all the missing subtraction terms. The repeated
procedure is somewhat lengthy and is thus presented in the Appendix
\ref{s:app} of this paper.

Note that the subtraction term from Equations
\ref{e:olness},\ref{e:olnessubt} can be combined with either of the
two hard cross-sections; joining it with (n-1)$^{\rm th}$
(Eq. \ref{e:olness0}) order cross-section instead of n$^{\rm th}$
(Eq. \ref{e:olness1}) one gets:
\begin{equation}
\bar{\sigma}^{(n-1)}_{\rm AB \to X} = (f_{H/A} - f_{g/A} \otimes
f^{(1)}_{H/g}) \otimes \hat{\sigma}^{(n-1)}_{\rm H s \to X} \otimes
f_{s/B}  = \bar{f}_{H/A} \otimes
\hat{\sigma}^{(n-1)}_{\rm H s \to X} \otimes f_{s/B}\, \, ,  
\label{e:olnessn-subt}
\end{equation}
where the new parton distribution function $\bar{f}_{H/A} =(f_{H/A} - f_{g/A} \otimes
f^{(1)}_{H/g})$ nicely expresses the physical origin of the subtraction term. 
In addition to being the collinear limit of the n$^{\rm th}$ order
calculation, it also represents the first (fixed) order component of
the QCD evolved parton distribution functions which is thus explicitly
removed from the fully re-summed (to all orders in $\alpha_s \ln(\mu^2/m^2_H)$) function $f_{H/A}$.

The experimentalists are generally interested in the fully
differential cross-sections from which exclusive topologies (events)
can be picked. The procedure to obtain an exclusive topology from an
inclusive differential cross-section is well established in various
flavors \cite{Schalicke:2005nv,Sjostrand:2006za,Corcella:2002jc} and is commonly known as the
Sudakov parton showering. Using the case of Eq. \ref{e:olness0} as an
example, the Factorization Theorem states that the incoming heavy quark
is fully resolved at the factorization scale $\mu=\mu_F$. The
probability $d{\rm P}_{g \to H \bar{H}}(\mu)$ of the heavy quark
un-resolving back to the gluon (back-evolution) at a
lower scale $\mu  < \mu_F $ (and producing an exclusive state with an additional
$\bar{H}$) is then given by the differential of the Sudakov
exponent (see \eg \cite{Sjostrand:2006za}):
\begin{eqnarray}
S_{c} &=& \exp{\left\{- \int\limits_{\mu^2}^{\mu_F^2} \frac{d\mu'^2}{\mu'^2} 
\frac{\alpha_s(\mu'^2)}{2\pi} \times \sum\limits_{a} 
\int\limits_{\xi_c}^{1} \frac{dz}{z}  P_{a \to c}(z) 
\frac{f_{a/I}(\frac{\xi_c}{z},\mu'^2)}{f_{c/I}(\xi_c,\mu'^2)}
\right\}}.
\label{e:sudakov_gen}\\
dS_{c}(\mu) &=& \sum\limits_{a} dS_{a \to c}(\mu) = \sum\limits_{a} \frac{d\mu^2}{\mu^2} 
\frac{\alpha_s(\mu^2)}{2\pi} \,  \frac{dz}{z}  P_{a \to c}(z) 
\frac{f_{a/I}(\frac{\xi_c}{z},\mu^2)}{f_{c/I}(\xi_c,\mu^2)} \times
S_{c} \notag  
\end{eqnarray}
For the given example of gluon splitting to heavy quarks
with $d{\rm P}_{g \to H \bar{H}}(\mu)\, = dS_{g \to H}(\mu)\,$ one gets:
\begin{equation}
d\sigma^{\rm shower}_{\rm AB \to X \bar{H} } = dS_{g \to H}(\mu)\, f_{H/A}(\mu_F)\, d\hat{\sigma}^{(n-1)}_{\rm H s \to X}\, f_{s/B}(\mu_F)\,\, ,  
\label{e:dshowern}
\end{equation}
which is also written as fully differential in all variables. Note
that the $dS_{g \to H}(\mu)$ term in the Monte-Carlo context
represents an explicit showering step (\ie a parton-showered parton
level 'hard' event).  A fact also
worth noting is that this expression is now of the order n and
contains one resolved particle $\bar{H}$, thus giving the same
configuration as the perturbative n$^{\rm th}$ order expression of
Eq. \ref{e:olness1} in the region ($\mu < \mu_F$). 
There is clearly an overlap between the
results of the two in this region, producing the `double-counting' in the overlap
region. The subtraction term of Eq. \ref{e:olnessubt} however retains
its role also in differential form:
\begin{equation}
d\sigma^{(n)}_{\rm subt} =  df^{(1)}_{H/g}(\mu)\, d\hat{\sigma}^{(n-1)}_{\rm
 H s \to X } \,  ~~~~~~(\mu < \mu_F) 
\label{e:dolnessubt}
\end{equation}

and this fully differential subtraction term depends on the variable
$\mu$ through:
\begin{equation}
df^{(1)}_{H/g} (\mu)\, = \frac{\alpha_s(\mu_F)\,}{2\pi} 
P_{g \to H}(z) \frac{dz}{z}  \frac{d\mu^2}{\mu^2},
\end{equation}
where the $\mu$ variable needs to be explicitly kinematically related
to the mass singularity in the (differential) n$^{\rm th}$ order
cross-section. The Equation \ref{e:olness1} can now be rewritten in
differential form:
\begin{equation}
d\sigma^{(n)}_{\rm AB \to X \bar{H}} = f_{g/A} \, d\hat{\sigma}^{(n)}_{\rm  g s \to X \bar{H} } \, f_{s/B}\, = 
f_{g/A} \, d\sigma^{(n)}_{\rm  g s \to X \bar{H} } \, f_{s/B}\, 
- f_{g/A} \, d\sigma^{(n)}_{\rm subt} \, f_{s/B}\, .  
\label{e:dolness1}
\end{equation}

Correspondingly, anticipating the exclusive final state
containing one heavy quark, the differential form of Equation \ref{e:olness} is then given by:
\begin{eqnarray}
d\sigma_{\rm AB \to X \bar{H}}  &=&  
f_{g/A} \, d\sigma^{(n)}_{\rm  g s \to X \bar{H} } \, f_{s/B}\, \label{e:dolness} \\
&-& f_{g/A} \, d\sigma^{(n)}_{\rm subt } \, f_{s/B}\,
\notag \\  
&+& d\sigma^{\rm shower}_{\rm AB \to X \bar{H} } \,, ~~~~~~~~~~~(\mu < \mu_F)\notag  
\end{eqnarray}
which, when inserting the explicit expressions from Equations  \ref{e:dshowern} and \ref{e:dolnessubt} gives:
\begin{eqnarray}
d\sigma_{\rm AB \to X \bar{H}}  &=&  
f_{g/A}(\mu_F) \, d\sigma^{(n)}_{\rm  g s \to X \bar{H} } \,
f_{s/B}(\mu_F)\, \label{e:dolnesse} \\
&-&  f_{g/A}(\mu_F)\, df^{(1)}_{H/g}(\mu)\, d\hat{\sigma}^{(n-1)}_{\rm
 H s \to X }\,  f_{s/B}(\mu_F)\,  ~~~~~~~~~~(\mu < \mu_F) \notag \\
&+& dS_{g \to H}(\mu)\, f_{H/A}(\mu_F)\, d\hat{\sigma}^{(n-1)}_{\rm H s \to X}\, f_{s/B}(\mu_F)\,\, .  ~~~~~~(\mu < \mu_F) \notag
\end{eqnarray}

It is trivially obvious that the subtraction term still cancels the
collinear limit (mass divergence) of the perturbative n$^{\rm th}$
order calculation of Eq. \ref{e:dolness1}, while its impact on the
(n-1)$^{\rm th}$ order (showered) calculation is again best seen by
combining the last two lines of Equation \ref{e:dolnesse} and results of Eqns. \ref{e:dshowern} and \ref{e:dolnessubt} into:
\begin{equation}
d\bar{\sigma}^{(n-1)}_{\rm AB \to X \bar{H}} = \left( dS_{g \to H}(\mu)\, f_{H/A}(\mu_F)\, - f_{g/A}(\mu_F)\,
df^{(1)}_{H/g}(\mu)\, \right) d\hat{\sigma}^{(n-1)}_{\rm H s \to X}\,  f_{s/B}(\mu_F)\,  \, .  
\label{e:dshowern-subt}
\end{equation}
Taking the limit $\mu \to \mu_F$ one quickly sees that:
\begin{eqnarray}
dS_{g \to H}(\mu)\, f_{H/A}(\mu_F)\, 
&\underrightarrow{\mu \to \mu_F}&
f_{g/A}(\mu_F)\, \frac{\alpha_s(\mu_F)\,}{2\pi} 
P_{g \to H}\, d\Phi \\
f_{g/A}(\mu_F)\, df^{(1)}_{H/g}(\mu)\,  
&\underrightarrow{\mu \to \mu_F}&
f_{g/A}(\mu_F)\, \frac{\alpha_s(\mu_F)\,}{2\pi} 
P_{g \to H}\, d\Phi,
\end{eqnarray}
with $d\Phi$ denoting all the variables in the differential. In this
limit $\mu \to \mu_F$ the two terms thus cancel \emph{on paper},
\ie exactly. Since $\mu_F$ is by definition the highest
virtuality reachable by parton showering approach, the kinematic region $\mu >
\mu_F$ is thus populated solely by the n$^{\rm th}$ order contribution and
the continuation on the transition point is \emph{smooth}. Choosing a
kinematic relation which defines $\mu$ in terms of kinematic
quantities measuring the collinearity (\eg transverse momentum of the
splitting, virtuality of the participating particle/propagator) one
achieves an ordered subtraction scheme ranging from the collinear
region to the showering limit $\mu_F$.

The result of these deliberations is that the subtraction term besides
it's obvious role actually  \emph{smoothly interpolates} between the
parton-shower evolved (n-1)$^{\rm th}$ order and n$^{\rm th}$ order perturbative
expressions while removing the double-counting contributions in the
overlap region ($\mu \leq \mu_F$).

There is no obvious reason why this procedure cannot be made iterative
since  the subtraction terms are designed to cancel all
mass divergences present in a perturbative calculation of a certain
order. In fact, a consistent iterative (and recursive) procedure has
been developed and implemented in our paper \cite{Kersevan:2006fq}, however the
iterative nature of it was not tested in the examples developed thus
far. The process of the associated Z boson production with two heavy
(b) quarks $gg \to Z^0/\gamma b \bar{b}$ serves as a good test case
when one wants both b-quarks to be resolved (observable) and correctly
described over the whole phase space. 

\subsection{The Implementation of the  Showering and Overlap Removal\label{s:implementation} }

In order to implement the above procedure in a Monte Carlo algorithm
one needs to define a specific mapping of the variables $\mu,z$ in the
Sudakov (parton) showering algorithm (c.f. Equation \ref{e:dshowern}) as well as appropriate kinematic
transforms between the four-momenta of the partons undergoing the
showering. 

The choice of the kinematic mappings and transforms to suit our
procedure is in principle arbitrary, as long as it properly takes into
account the heavy parton masses. Consequently the defined procedure
could be matched to the parton showers of e.g. Pythia
\cite{Sjostrand:2006za} or Herwig \cite{Corcella:2002jc} if the mass
of the partons were incorporated.

Nevertheless, we chose to implement our own showering algorithm based
on the one suggested by Collins \emph{et al}
\cite{Chen:2001ci,Collins:2002ey,Collins:2000qd,Chen:2001nf} and
extended it to properly incorporate heavy parton masses. The explicit
derivation of the showering algorithm, kinematic mappings and
transforms applied to our procedure have already been described in our
previous paper \cite{Kersevan:2006fq}. Summarized briefly, the
implemented showering algorithm has the evolution variable $\mu$ 
chosen to be equal to the virtuality of the particle (in our case
heavy quark) and in each showering step the rapidity of the showered
system is preserved, which sets the variable $z$ of Equation
\ref{e:sudakov_gen}. The value of the factorization scale $\mu_F$ is
not pre-determined (\ie can be picked from one of reasonable options,
\eg the invariant mass of the subsystem or similar).

The choice of the showering algorithm implementation was motivated by two main points:
\begin{itemize}
\item This subtraction term of Equation \ref{e:dolnessubt}, albeit
  derived in a different way, is in form identical to the subtraction term
  derived by Collins \emph{et al}. This fact motivated us to implement the parton
  showering algorithm according to prescriptions given in the cited
  papers.
\item The selected kinematic setup is motivated by the fact that, as
  shown by Collins \emph{et al}
  \cite{Chen:2001ci,Collins:2002ey,Collins:2000qd,Chen:2001nf}, the
  procedure of parton showering and subtraction is not equivalent to
  the standard subtraction schemes (\eg $\rm \overline{MS}$) and thus
  the parton distribution functions should in principle be
  modified. The modification is generally non-trivial (using
  e.g. Pythia \cite{Sjostrand:2006za} showers), however using
  this specific choice of showering kinematics the corrected ($\rm JCC$) scheme
  is for quarks simply related to the $\rm \overline{MS}$ parton
  distribution functions:
\begin{eqnarray}
z\,f^{\mathrm{JCC}}_{i/I}(z,\mu^2) &=& z\,f^{\mathrm{\overline{MS}}}_{i/I}(z,\mu^2) \\
\notag &+& \frac{\alpha_s(\mu^2)}{2\pi} \int\limits_z^1 \,d\xi \frac{z}{\xi} 
  f^{\mathrm{\overline{MS}}}_{g/I}(\xi,\mu^2) 
\left[ P_{g \to i}(\frac{z}{\xi}) \ln\left(1 -  \frac{z}{\xi}\right) +
\frac{z}{\xi}\left(1 -  \frac{z}{\xi}\right) \right] \\ \notag
&+& \mathcal{O}\mathrm{(\text{first-order quark
terms})}~+\mathcal{O}(\alpha_s^2)\, .
\label{e:jcc}
\end{eqnarray}
  The above Equation \ref{e:jcc} does not apply to gluons. It needs to
  be pointed out that at present it is not known how to simply extend
  this approach to gluons. For the interested reader on the work in
  this direction we recommend the paper \cite{Collins:2004vq}.
\
\end{itemize}

It needs to be emphasised that for the purpose of this paper, only a
very narrow implementation of the showering algorithm was needed,
namely a single heavy quark backward branching to gluon and subsequently only
this part was actually implemented in place of a full parton showering
program.

We can now summarize the properties of thus obtained algorithm of
combining pQCD and parton showering as implemented in our Monte-Carlo algorithm:
\begin{itemize}
\item All heavy quarks (incoming and outgoing) are kept \emph{massive}
  throughout the procedure, both in the perturbative calculation of
  matrix elements and in the showering procedure and overlap
  removal. The matrix elements used are at present `leading order' (tree-level)
  only, however the procedure could in principle be expanded to
  diagrams containing virtual corrections.
\item All overlap removal is done at the parton level on an
  event-by-event basis. The collinear topologies are determined from
  the participating Feynman diagrams (currently done manually but
  could in theory be automatized).
\item The kinematics of the shower and overlap removal is implemented
  especially for this approach. These choices are reflected in the
  subtraction term which is achieved by calculating the collinear
  limit of the kinematic topology (event) of the n$^{\rm th}$
  order perturbative calculation.
\end{itemize}

The event generation is thus performed in the following steps:
\begin{itemize}
\item The n$^{\rm th}$ order process (event) is sampled, the
  collinear limits for the given event topology are estimated and the
  subtraction terms of order (n-1) are calculated. The weight is given
  by Equation \ref{e:dolness1}.
\item The (n-1)$^{\rm th}$ order process (event) is sampled. If this
  process still contains gluon splitting to heavy quarks in the
  initial state (in the other incoming leg) the corresponding
  subtraction terms of order (n-2) are again calculated. The event is
  then showered and the weight is given by Equation \ref{e:dshowern}.
\item If subtraction terms were found in the previous step, the
  corresponding (n-2)$^{\rm th}$ order process is calculated and showered on both
  legs to achieve the (n$^{\rm th}$ order) event configuration of the
  previous two cases. The weight is estimated analogous to Equation
  \ref{e:dshowern} for two showering steps.
\end{itemize}
All the above classes of events are separately unweighted, obtaining
weights equal to $\pm 1$ because in some phase space points the
contributions from subtraction terms can actually be greater than the
unsubtracted values, as will be shown in the following sections. The
summed contributions of all processes, corresponding to Equation
\ref{e:dolness}, are of course positive throughout the phase space.  In
the actual Monte-Carlo program, it is very easy to (pre-)mix these
classes of events internally to obtain the summed contribution
corresponding to Equation \ref{e:dolness}.

The procedure of \cite{Kersevan:2006fq} has
been improved by introducing the \emph{massive} splitting kernel
correction for the initial state $g \to H \bar{H}$ splits, which can
be derived from \cite{Catani:2002hc} results:
\begin{equation}
P_{g \to H}^{\rm massive} = P_{g \to H}^{\rm massless} + P_{g \to
H}^{\rm correction} = {\rm T_R}\, \left( 1 - 2z(1-z) \right) + {\rm T_R}\,\left(
\frac{2z(1-z) m_H^2}{p_T^2 + m_H^2} \right),
\label{e:pmassive}
\end{equation}
with $p_T$ being the relative transverse momentum of the spectator
heavy quark in the split. The massive kernels have already been used for
final state showering in the Sherpa Monte-Carlo generator \cite{Schalicke:2005nv} but have to our
knowledge never been used in the showering of initial state heavy
quarks.
 
\subsection{The \zbbd Implementation}

Assuming an experimentalist at LHC wants to study the Drell-Yan production
with two resolved (heavy) b-quarks in the final state, one now has three
possible choices of calculation (c.f. Figure \ref{f:zbbchoices}):
\FIGURE{
     \epsfig{file=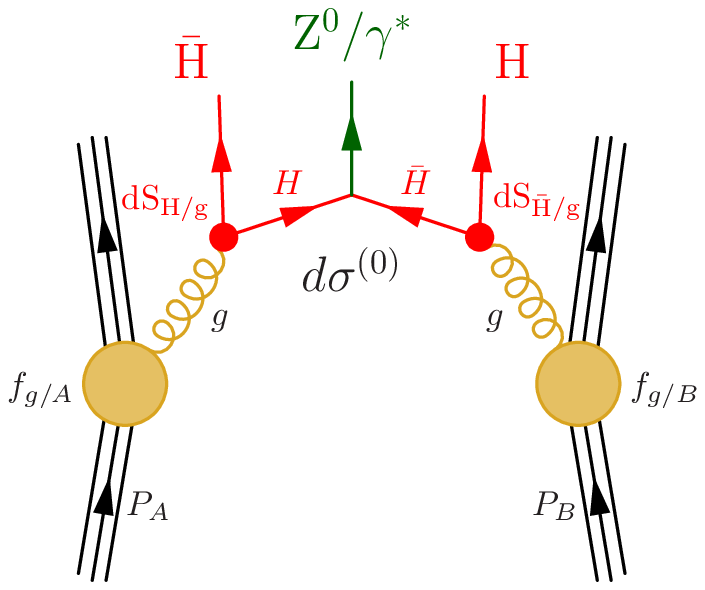,width=4.5cm}\parbox{0.3cm}{\vskip -3.5cm \Large$\oplus$}
     \epsfig{file=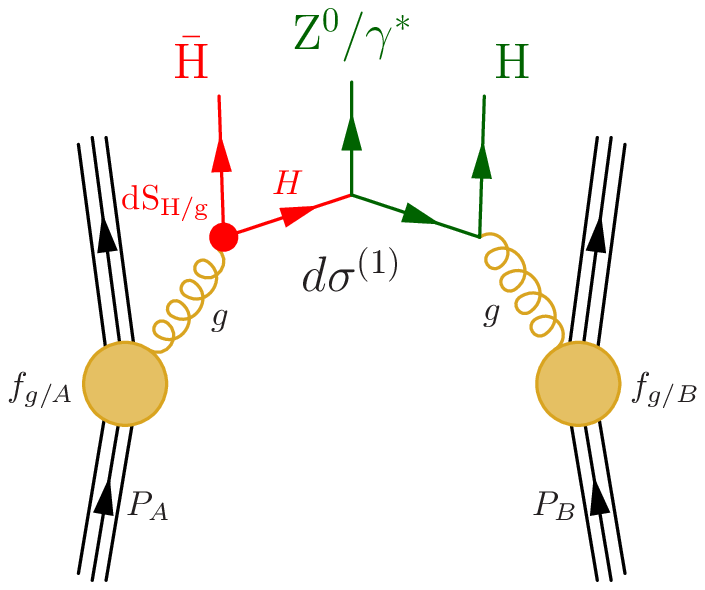,width=4.5cm}\parbox{0.3cm}{\vskip -3.5cm \Large$\oplus$}
     \epsfig{file=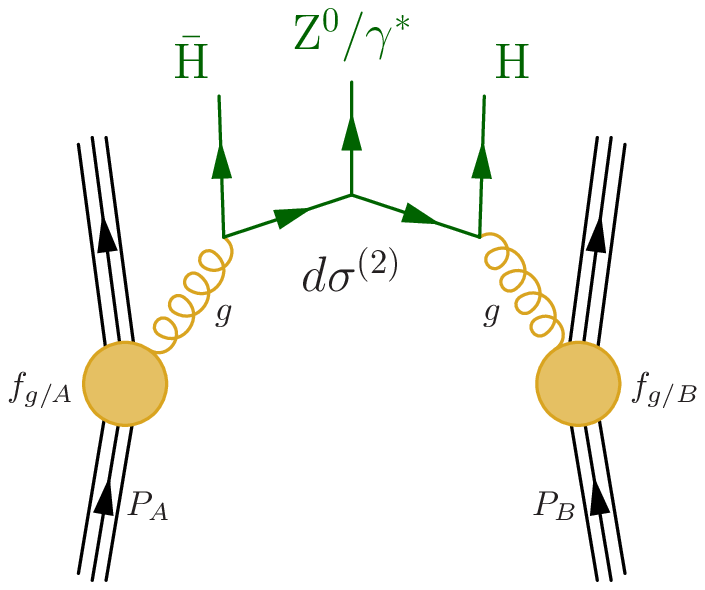,width=4.5cm}
\caption{ Schematic representation of contributions resulting in
exclusive $Z^0 H \bar{H}$ final state:  two fully evolved heavy (b)
quarks entering `pure' Drell-Yan at order $\alpha_s^0$ in combination
with double initial state parton shower (left), one heavy quark and
one gluon entering the hard process at order $\alpha_s^1$ in combination with one parton
shower (middle) and fully perturbative calculation involving two
incoming gluons in a hard process of order $\alpha_s^2$ (right). These
three processes need to be combined with appropriate overlap removal
as detailed in the text. \label{f:zbbchoices}}
}

\begin{itemize}
\item The order $\alpha_s^0$ hard process $b \bar{b} \to Z^0/\gamma^*
\to f \bar f$ with fully evolved b-quarks entering the hard process at
$\mu_F$. The cross-section contains no mass singularities and needs no
subtraction terms ($d\hat{\sigma}^{(0)} = d\sigma^{(0)}$). The
associated  b-quarks are then resolved at scales
$\mu_{1H}$ and $\mu_{2\bar{H}}$ using parton showering:
\begin{equation}
d{\sigma}^0_{\zbbp} = \sum\limits_{H=b,\bar{b}}
 dS_{g \to H}(\mu_{1H})\, f_{H/A}(\mu_F)\, d{\sigma}^{(0)}_{\rm H \bar{H}
\to \zz }\, dS_{g \to \bar{H}}(\mu_{2\bar{H}})\, f_{\bar{H}/B}(\mu_F)\,\, .  
\label{e:dzbb0}
\end{equation}
Kinematically, the leg with higher induced virtuality (scale) is
treated (unresolved to gluon) first. 
\item The order $\alpha_s^1$ hard process $g H \to \zdec H, \,
H=b,\bar{b}$ with one fully evolved b-quark entering the hard process at
$\mu_F$ and the other one participating as the propagator inside the
matrix element calculation. The cross-section thus contains one mass
singularity related to the propagator and needs a subtraction term
derived from the collinear limit of the b-quark propagator. The other
associated b-quark is then resolved using parton showering and the
scales $\mu_{1,2}$ are set to be the evolution scale of the showered
quark and the virtuality of the b-quark propagator respectively:
\begin{eqnarray}
d{\sigma}^1_{\zbbp} &=& \sum\limits_{H=b,\bar{b}}
 dS_{g \to H}(\mu_{1H})\, f_{H/A}(\mu_F)\, d\sigma^{(1)}_{\rm H g
\to \zz H }\,  f_{g/B}(\mu_F)\,  \label{e:dzbb1} \\
&+& \sum\limits_{H=b,\bar{b}}  f_{g/A}(\mu_F)\, d\sigma^{(1)}_{\rm g \bar{H}
\to \zz \bar{H} }\, dS_{g \to \bar{H}}(\mu_{2\bar{H}})\, f_{\bar{H}/B}(\mu_F)\, \notag
 \\
&-& \sum\limits_{H=b,\bar{b}}  dS_{g \to H}(\mu_{1H})\, f_{H/A}(\mu_F)\,
d\sigma^{(0)}_{\rm  H \bar{H}
\to \zz }\, df^{(1)}_{\bar{H}/g}(\mu_{2\bar{H}})\, f_{g/B}(\mu_F)\, \notag
\\ 
&-& \sum\limits_{H=b,\bar{b}} f_{g/A}(\mu_F)\, df^{(1)}_{H/g}(\mu_{1H})\,
d\sigma^{(0)}_{\rm H \bar{H}
\to \zz }\, dS_{g \to \bar{H}}(\mu_{2\bar{H}})\, f_{\bar{H}/B}(\mu_F)\, \notag
\, .
\end{eqnarray}
\item The order $\alpha_s^2$ hard process \zbbd where both incoming
b-quarks participate as propagators inside the 
matrix element calculation. The cross-section thus contains two mass
singularities related to the propagators and needs
corresponding subtraction terms  (with permutations)  
derived from the collinear limit of the two b-quark propagators,
virtualities of which then define the
scales $\mu_{1,2}$. Using the formalism of \cite{Kersevan:2006fq}, or equivalently
\cite{Olness:1997yc}, one obtains:
\begin{eqnarray}
d{\sigma}^2_{\zbbp} &=&  f_{g/A}(\mu_F)\, d\sigma^{(2)}_{\zbb }\,
f_{g/B}(\mu_F)\, \label{e:dzbb2} \\
&-& \sum\limits_{H=b,\bar{b}} f_{g/A}(\mu_F)\, df^{(1)}_{H/g}(\mu_{1H})\,
d\sigma^{(1)}_{\rm H g \to \zz H }\,  f_{g/B}(\mu_F)\, \notag\\
&-& \sum\limits_{H=b,\bar{b}}  f_{g/A}(\mu_F)\, d\sigma^{(1)}_{\rm g \bar{H}
\to \zz \bar{H} }\,  df^{(1)}_{\bar{H}/g}(\mu_{2\bar{H}})\, f_{g/B}(\mu_F)\,
\notag \\
&+& \sum\limits_{H=b,\bar{b}} f_{g/A}(\mu_F)\, df^{(1)}_{H/g}(\mu_{1H})\,
d\sigma^{(0)}_{\rm H \bar{H} \to \zz }\,
df^{(1)}_{\bar{H}/g}(\mu_{2\bar{H}})\, f_{g/B}(\mu_F)\,\, .
\notag
\end{eqnarray}
\end{itemize}
The derivation of the above subtraction terms is presented in Appendix
\ref{s:app}.

One might be surprised at the $+$ sign of the last `subtraction' term
in Eq. \ref{e:dzbb2} however re-arranging the subtraction terms
to be matched with the corresponding showering expressions in the spirit of
Eq. \ref{e:dshowern-subt} one derives the tree cross-section differentials:
\begin{eqnarray}
d\bar{\sigma}^0 &=& \sum\limits_{H=b,\bar{b}}
 \left(dS_{g \to H}(\mu_{1H})\, f_{H/A}(\mu_F)\, - \,
df^{(1)}_{H/g}(\mu_{1H})\,f_{g/A}(\mu_F)\right) 
d\sigma^{(0)}_{\rm H \bar{H} \to \zz }\, \times \notag \\
&\times&  
\left(dS_{g \to \bar{H}}(\mu_{2\bar{H}})\, f_{\bar{H}/B}(\mu_F)\,- df^{(1)}_{\bar{H}/g}(\mu_{2\bar{H}})\, f_{g/B}(\mu_F )\right) \,   
\label{e:dzbb0bar}
\end{eqnarray}
\begin{eqnarray}
d\bar{\sigma}^1 &=& \sum\limits_{H=b,\bar{b}}
\left(dS_{g \to H}(\mu_{1H})\, f_{H/A}(\mu_F)\, - \,
df^{(1)}_{H/g}(\mu_{1H})\,f_{g/A}(\mu_F)\right)  d\sigma^{(1)}_{\rm H g
\to \zz H }\,  f_{g/B}(\mu_F)\,   \notag \\
&+& \sum\limits_{H=b,\bar{b}}  f_{g/A}(\mu_F)\, d\sigma^{(1)}_{\rm g \bar{H}
\to \zz \bar{H} }\, \left(dS_{g \to \bar{H}}(\mu_{2\bar{H}})\,
f_{\bar{H}/B}(\mu_F)\, -  df^{(1)}_{\bar{H}/g}(\mu_{2\bar{H}})\, f_{g/B}(\mu_F
)\right) \notag \\ \label{e:dzbb1bar} \, 
\end{eqnarray}
and trivially:
\begin{equation}
d\bar{\sigma}^2 =  f_{g/A}(\mu_F)\, d\sigma^{(2)}_{\zbb }\,
f_{g/B}(\mu_F)\, \label{e:dzbb2bar}.
\end{equation}

\FIGURE{
     \epsfig{file=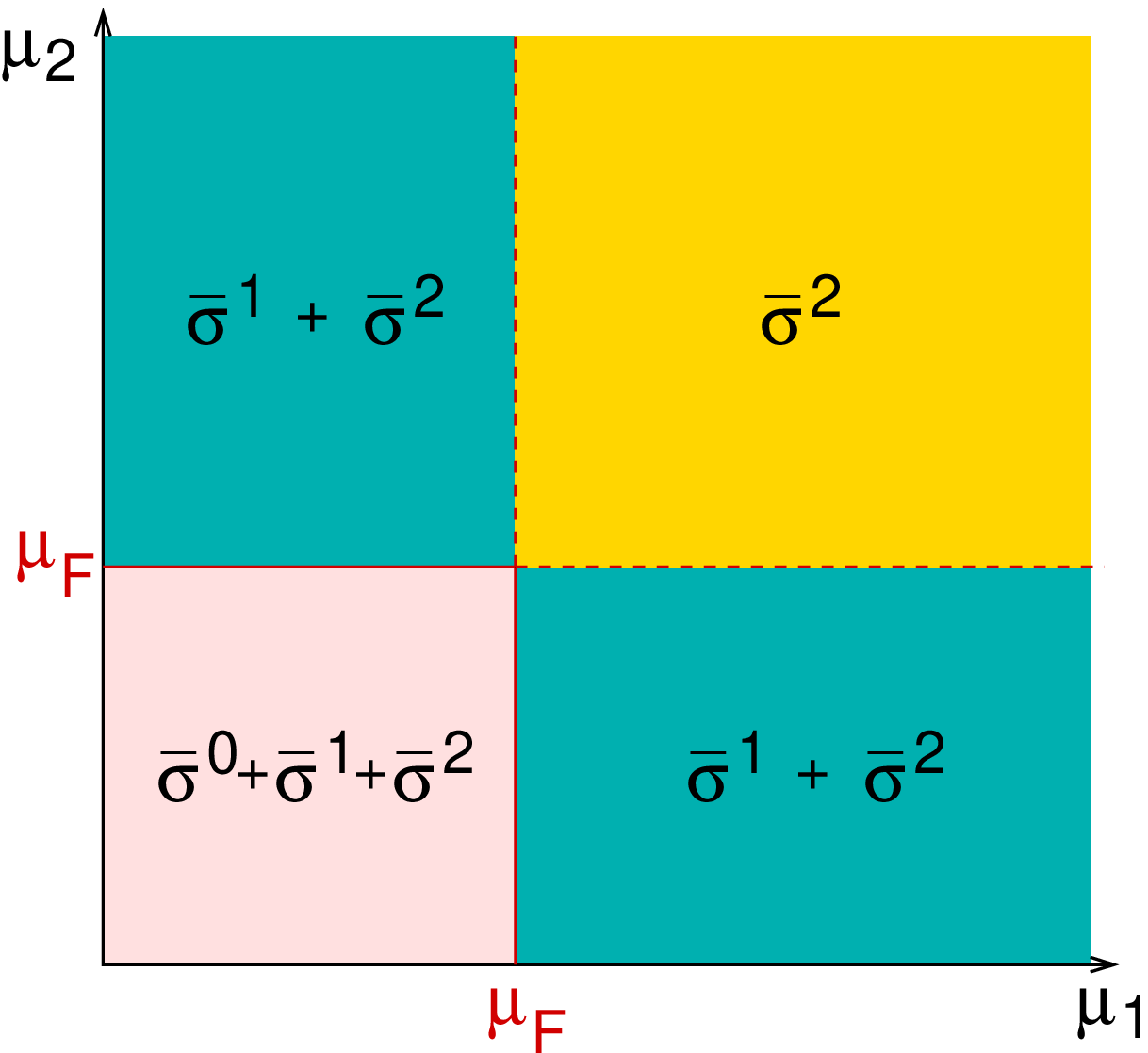,width=6.2cm}
\caption{ Schematic representation of contributions sources
corresponding to four scenaria 
$\mu_{1,2} < \mu_F$ where all the tree contributions are non-zero,
$\mu_1 < \mu_F, \mu_2 > \mu_F$ and  $\mu_2 < \mu_F, \mu_1 > \mu_F$
where only the contributions from Eq. \ref{e:dzbb1bar} and
\ref{e:dzbb2bar} are non-zero and the region $\mu_{1,2} > \mu_F$ where
only the fully perturbative contribution of Eq. \ref{e:dzbb2bar}
contributes.\label{f:muregions}}
}
As noted above, there are actually four scales $\mu_{1H,2\bar{H}},
(H=b,\bar{b})$, namely the virtualities assuming that the
b-quark (anti-quark) are connected to the first (second) gluon from
hadrons A and B.
From the above forms it is evident that the first contribution from
Eq. \ref{e:dzbb0bar} is zero when either of $\mu_{1,2} = \mu_F$ as are
respective terms in Eq. \ref{e:dzbb1bar}, ensuring the smooth
transitions of the functions in the region close to the factorization
scale as expected. Subsequently, the behavior of the subtraction
terms corresponds to four scenaria 
$\mu_{1,2} < \mu_F$ where all the tree contributions are non-zero,
$\mu_1 < \mu_F, \mu_2 > \mu_F$ and  $\mu_2 < \mu_F, \mu_1 > \mu_F$
where only the contributions from Eqns. \ref{e:dzbb1bar} and
\ref{e:dzbb2bar} are non-zero and the region $\mu_{1,2} > \mu_F$ where
only the fully perturbative contribution of Eq. \ref{e:dzbb2bar}
contributes. Graphically, the contributions are shown in Figure
\ref{f:muregions}.

At this point one should also review the omissions (limitations) of the proposed
algorithm: As presented in Equations \ref{e:dzbb0}-\ref{e:dzbb2bar}
the terms $dS_{g \to H}(\mu)$ indicate that there is no emission
between the scales $[\mu_{1,2},\mu_F]$ (c.f. Eq. \ref{e:dshowern}), \ie the
heavy quark unresolves (backward evolution) directly to gluon. All the
tree contributions will thus uniquely produce the same initial ( $\rm
g g$) and final ($\rm \zz b \bar{b}$) state.

In a full shower implementation as \eg in Pythia \cite{Sjostrand:2006za},
HERWIG \cite{Corcella:2002jc} or Sherpa \cite{Schalicke:2005nv}, one
or more additional branchings in terms of gluon radiation $\rm H \to g H$
could take place before the heavy quark would resolve back to a gluon
via gluon splitting. Consequently, a part of the contributions, which
would be present in case of a full shower from inclusive ${\rm g
  H \to \zz H }$ and ${\rm H \bar{H} \to \zz }$ production,
is missing. The probabilities of omitted showering contributions are
however considered to be small, using again the argument of the order
of heavy quark density $f_{H/I}$ being of the effective order
$\alpha_s$ higher with respect to the gluon PDFs and considering the
$dS$ form of Eq. \ref{e:dshowern}.
 
\section{Implementation and Results}

The process \zbb is implemented in the AcerMC Monte-Carlo generator
\cite{Kersevan:2002dd} using the (adapted) MadGraph
\cite{Maltoni:2002qb} matrix elements with full $Z^0/\gamma^*$
interference. Unweighted events corresponding to the three
sub-processes, given by Equations \ref{e:dzbb0}, \ref{e:dzbb1} and
\ref{e:dzbb2}, are generated with native AcerMC single heavy quark
backward branching to gluon \cite{Kersevan:2006fq}. Each $dS$
term in the Eq. \ref{e:dzbb0}, \ref{e:dzbb1} and
\ref{e:dzbb2} thus corresponds to an actual parton showering step in
the event generation. 

As already stated, only a very narrow implementation of the showering
algorithm was needed for the purpose of this paper, namely a single
heavy quark backward branching, and subsequently only this part was
actually implemented in place of a full parton showering algorithm.
The produced parton level events from AcerMC thus need to be passed to
Monte Carlo tools as Pythia \cite{Sjostrand:2006za} or HERWIG
\cite{Corcella:2002jc} for further showering and hadronization. 

The showering algorithms in AcerMC (implemented in the style of Pythia
veto sampling) can in principle easily be expanded to full shower with
both initial and final state radiation and all possible branchings but
for purposes of this paper we only needed that single showering step.
Although the implementation of the full showering in AcerMC would
guarantee consistence by using only one showering algorithm (whereas
now it needs to be extended with further Pythia or Herwig showers with
different showering algorithms ), the subsequent interface of the
partons to the fragmentation algorithms in Pythia or Herwig would be
rather difficult and beyond the scope of this project.

Due to the subtraction terms a fraction of event candidates achieve
negative sampling weights and unweighted events are produced with
weight values of $\pm 1$ using the standard unweighting procedures. In
the subsequent studies only the pure $Z^0 \to \mu^+ \mu^- $ channels
were used for benchmarking (the photon contribution was turned off) in
the LHC environment (proton-proton collisions at $\rm \sqrt{s}$ = 14
TeV) in combination with the CTEQ6L1 \cite{Pumplin:2002vw} and hence
derived JCC PDF sets\footnote{There is a possibly valid argument that
  the NLO PDF sets should be used instead of LO ones since the derived
  procedure is in part NLO. The choice of these does not affect our
  results qualitatively and affects above all the absolute
  normalization (cross-section) prediction which is not correct anyway
  due to the absence of virtual corrections to our calculations.}  to
manifestly see the impact of using the JCC (Eq. \ref{e:jcc})  modification, which turns
out to be sizable in the b-quark case, contrary to its impact on the
light quark PDF values.  Let us emphasize again that Collins \emph{et
  al} have shown in
\cite{Chen:2001ci,Collins:2002ey,Collins:2000qd,Chen:2001nf} that the
PDF sets obtained e.g. in the $\overline{\rm MS}$ subtraction scheme as
the CTEQ sets are not formally the correct ones to be used in parton
showering but that instead modified PDF sets corresponding to the
showering algorithm (as the JCC for the Collins-style shower
prescription implemented in AcerMC)  should be used. In other
words, the PDFs in Monte-Carlo event generators are determined by the
showering algorithm and cannot be freely chosen, unlike the case for
PDFs used in inclusive calculations. 

The b-quark mass is set to $m_b = 4.8$ GeV and the factorization and
renormalization scales were chosen to be equal to the $Z^0$ mass
($\mu_F=M_Z$) but of course other choices are possible. All the
presented results are at parton level (i.e. b-quarks, $\zz$ and its
decay products) as output by AcerMC.

It is instructive to first reproduce the order $\alpha_s^1$ results
derived in our paper \cite{Kersevan:2006fq} in order to check the impact of the
massive evolution kernels introduced in this paper
(Eq. \ref{e:pmassive}). The results are presented in Figure
\ref{f:alpha1}. 
\FIGURE{
     \epsfig{file=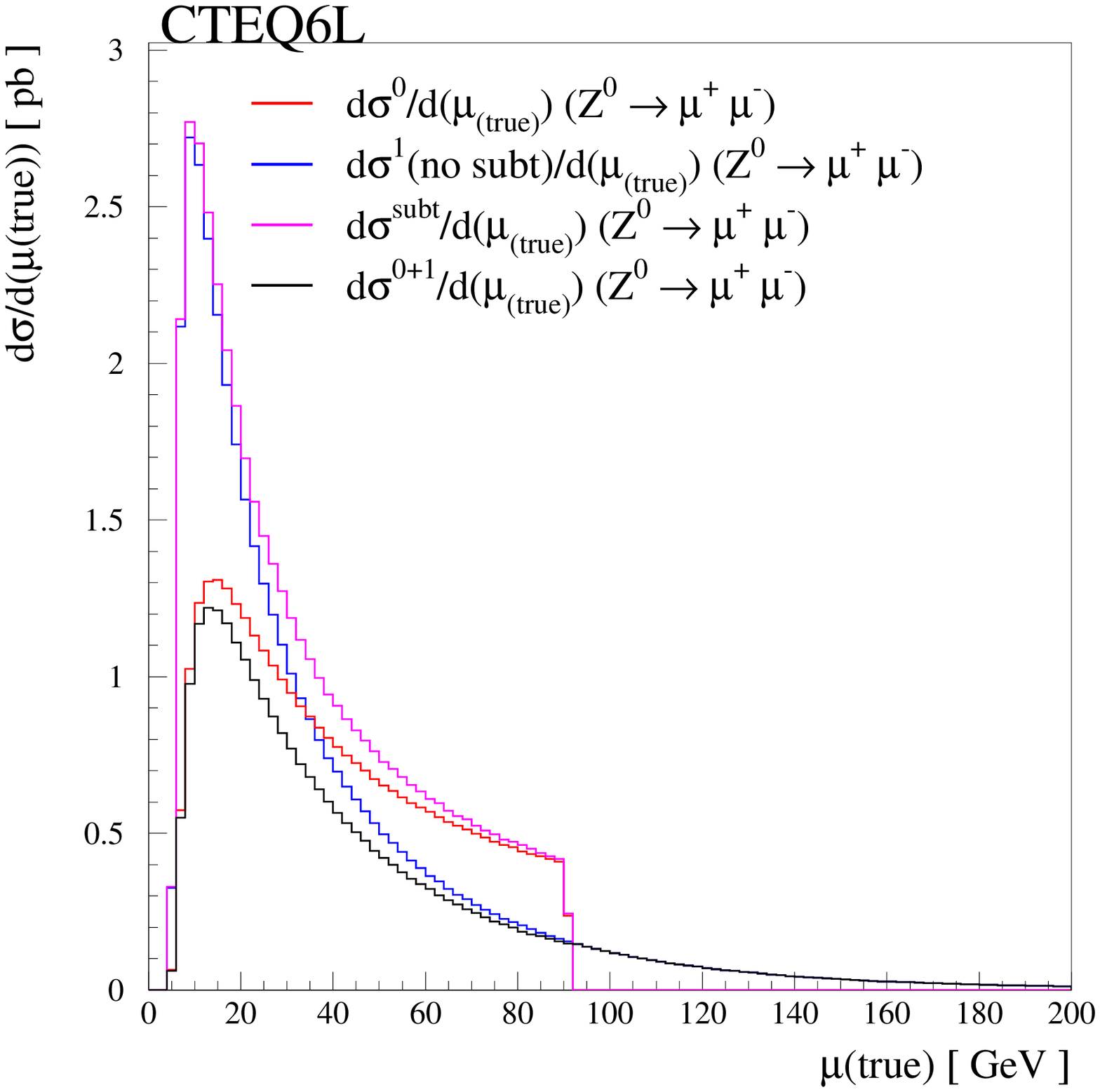,width=5.4cm}\hspace{-0.8cm}
     \epsfig{file=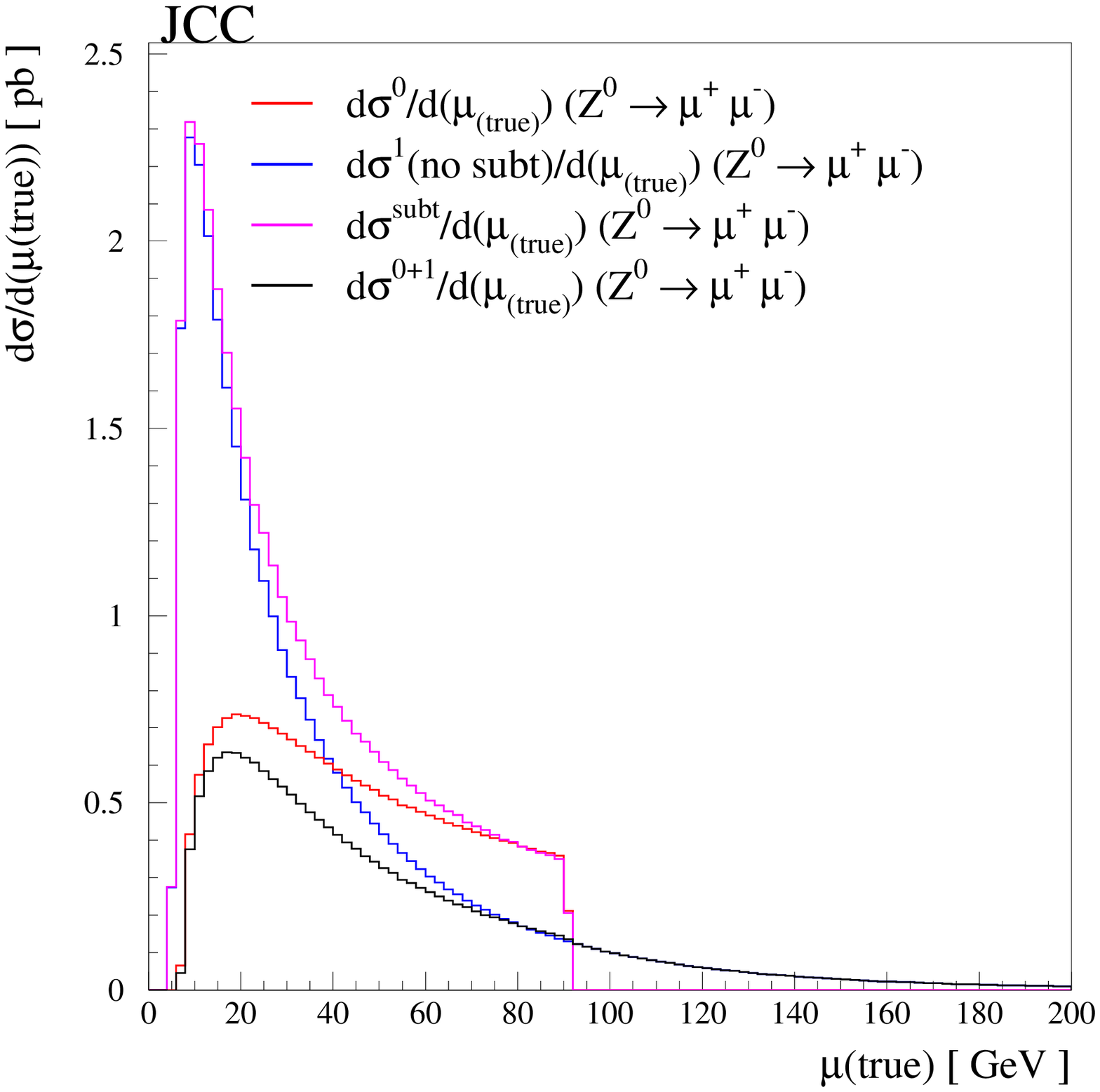,width=5.4cm}\hspace{-0.8cm}
     \epsfig{file=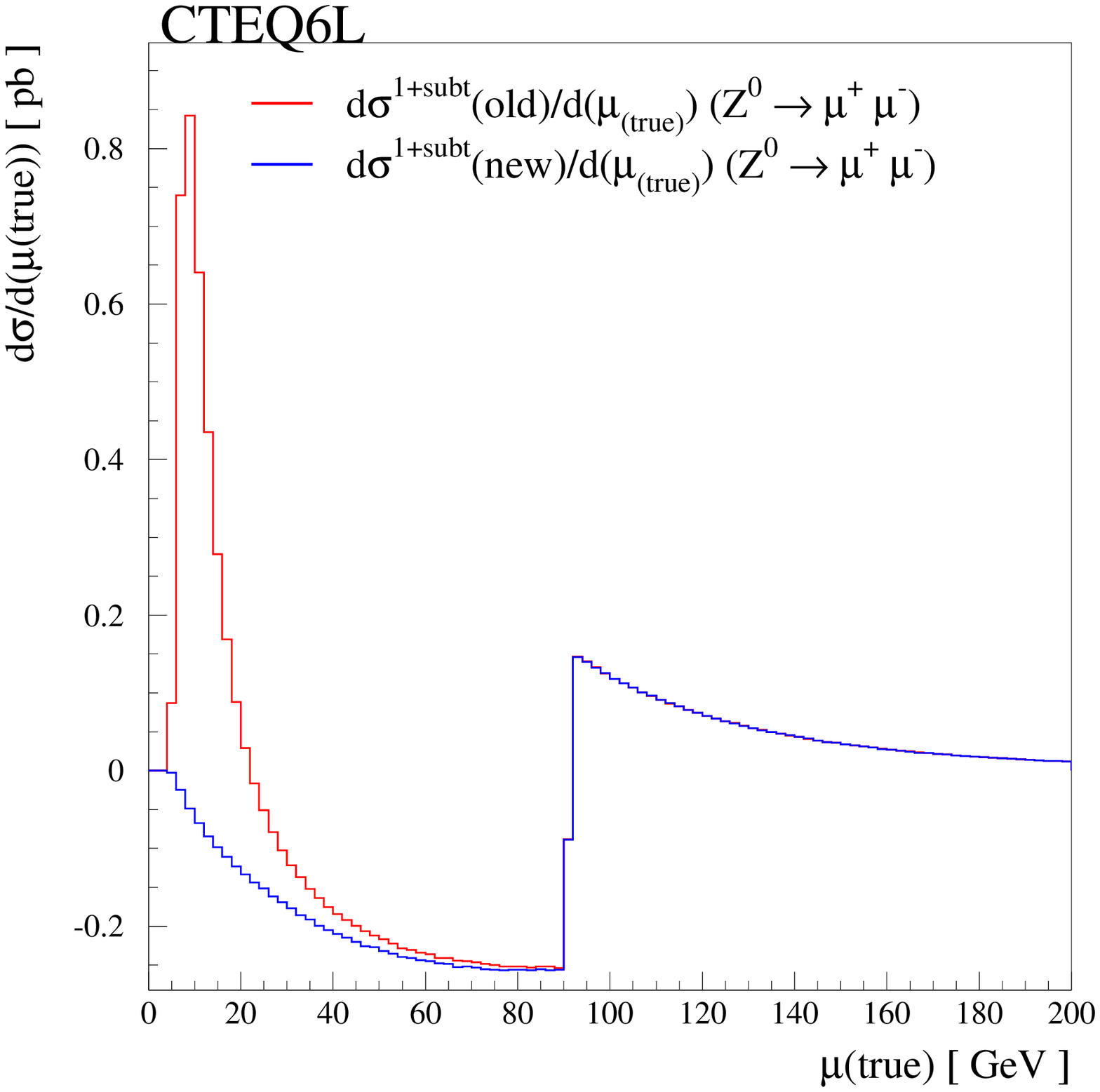,width=5.4cm}
\caption{ The reproduced order $\alpha_s^1$ results from \cite{Kersevan:2006fq}
using the CTEQ6L1 and JCC PDF sets (left and middle); note that the
subtraction term (magenta line) indeed gives a smooth transition from
low $\mu$ region, where it matches the perturbative calculation (blue)
to the $\mu=\mu_F$ region where it matches the parton shower
prediction (red). In the right plot the impact of introducing massive
splitting kernels is shown explicitly for the order $\alpha_s^1$
prediction combined with the subtraction term; note that the low $\mu$
peak previously present in \cite{Kersevan:2002dd} now disappears completely.  
\label{f:alpha1}}
}

As one can observe in the rightmost plot of Fig. \ref{f:alpha1} the
introduction of massive splitting kernels has corrected the low scale
(virtuality) region; the subtraction term now indeed smoothly
interpolates between the low scale (collinear limit) where it matches
the order $\alpha_s^1$ contribution to the factorization scale $\mu_F=
M_Z$ where it coincides with the parton shower prediction as expected,
thus allowing smooth transition between the two contributions as predicted. 
Note that at the order $\alpha_s^1$ \emph{two} virtualities for incoming
b-quarks are picked from
the Sudakov back-evolution (to gluon) of the order $\alpha_s^0$ pure
Drell-Yan process but only \emph{one} leg is actually
showered and correspondingly only one subtraction term is introduced
in the order $\alpha_s^1$ process. Since in this case there is no
ambiguity one can trace the actual $\mu$ used in the shower evolution and
subtraction (labeled $\mu(\rm true)$ in Figure \ref{f:alpha1}). 

\FIGURE{
     \hspace{0.7cm}\epsfig{file=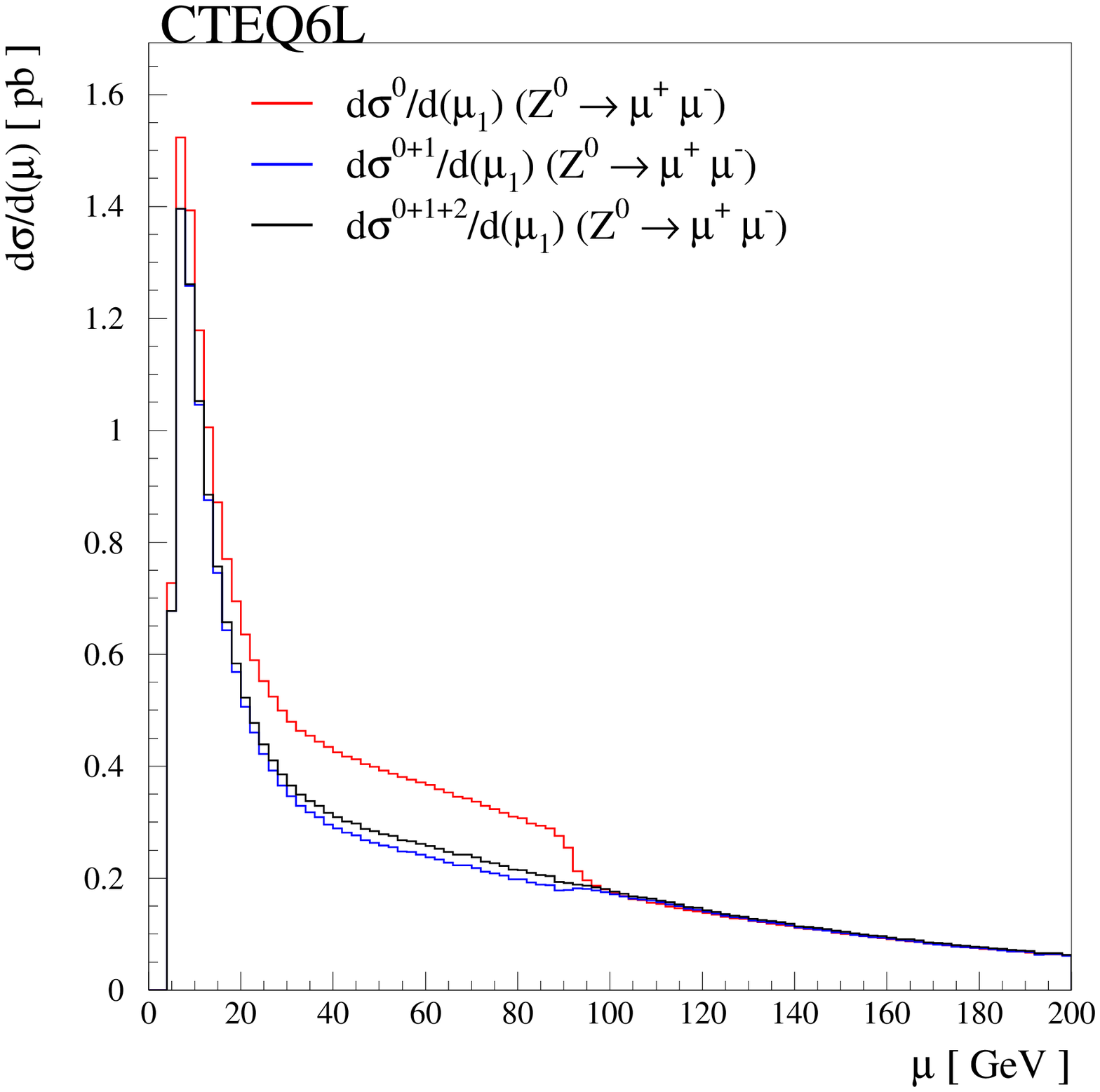,width=5.5cm}
     \epsfig{file=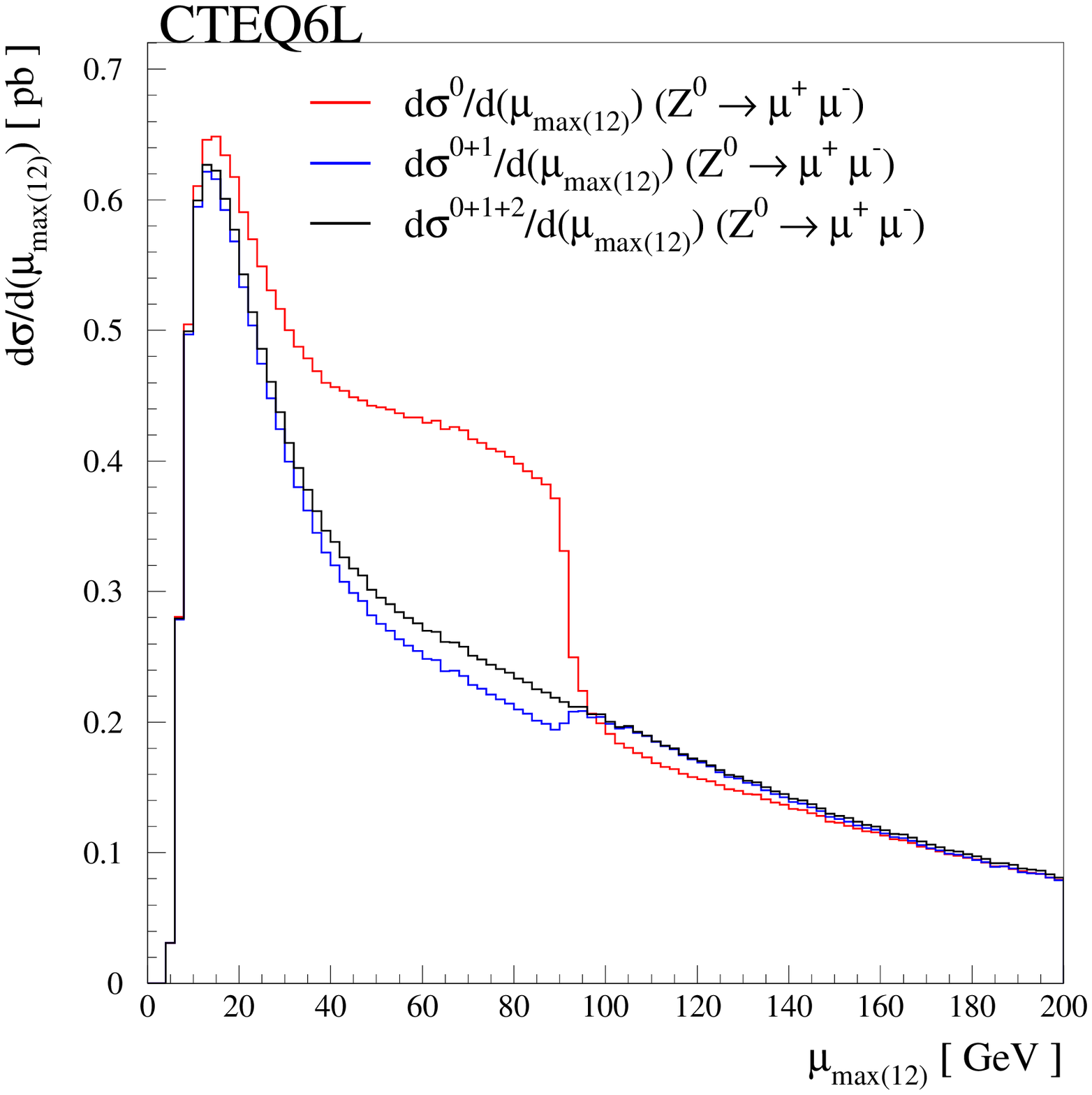,width=5.5cm}\\
     \epsfig{file=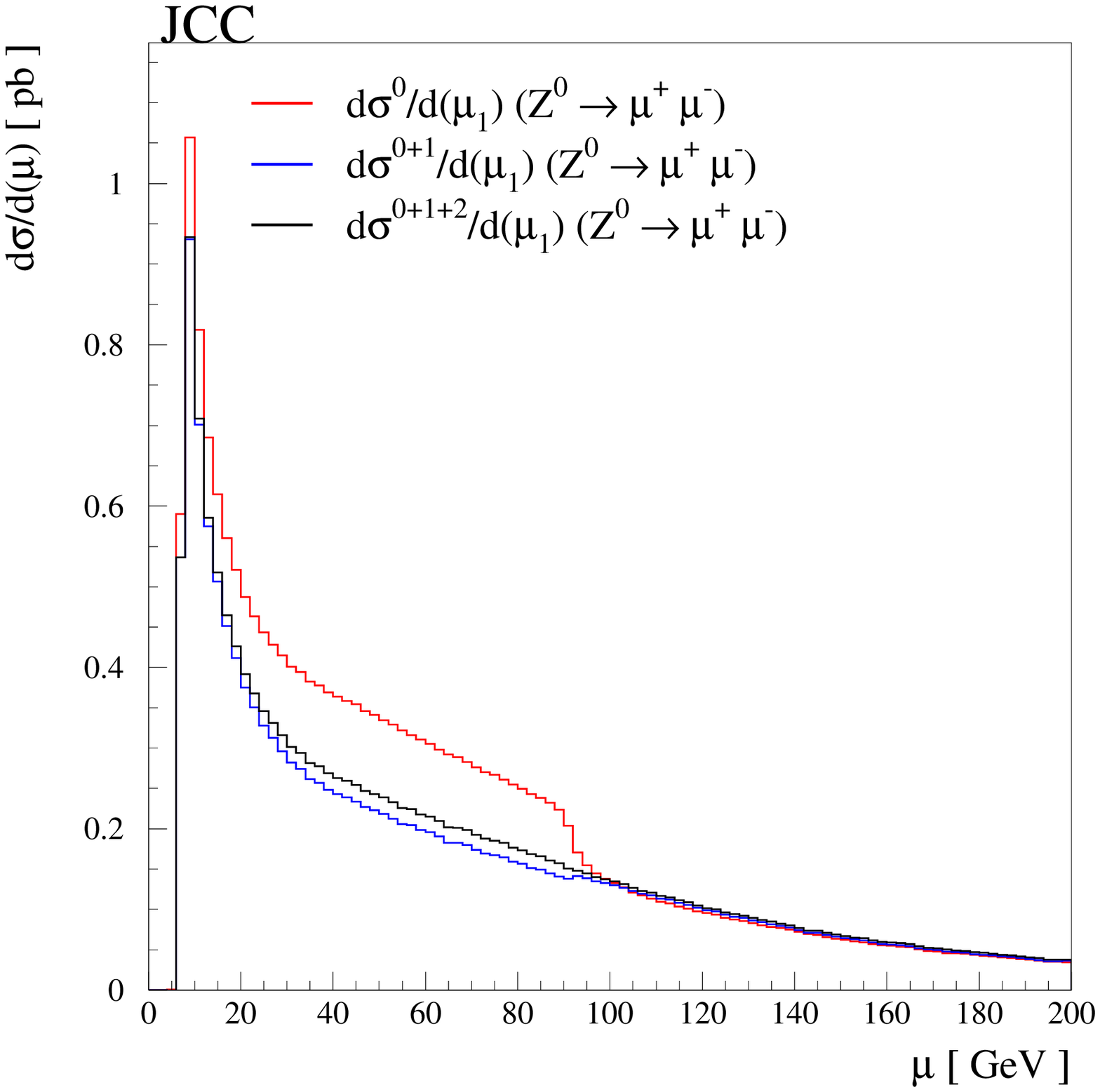,width=5.5cm}
     \epsfig{file=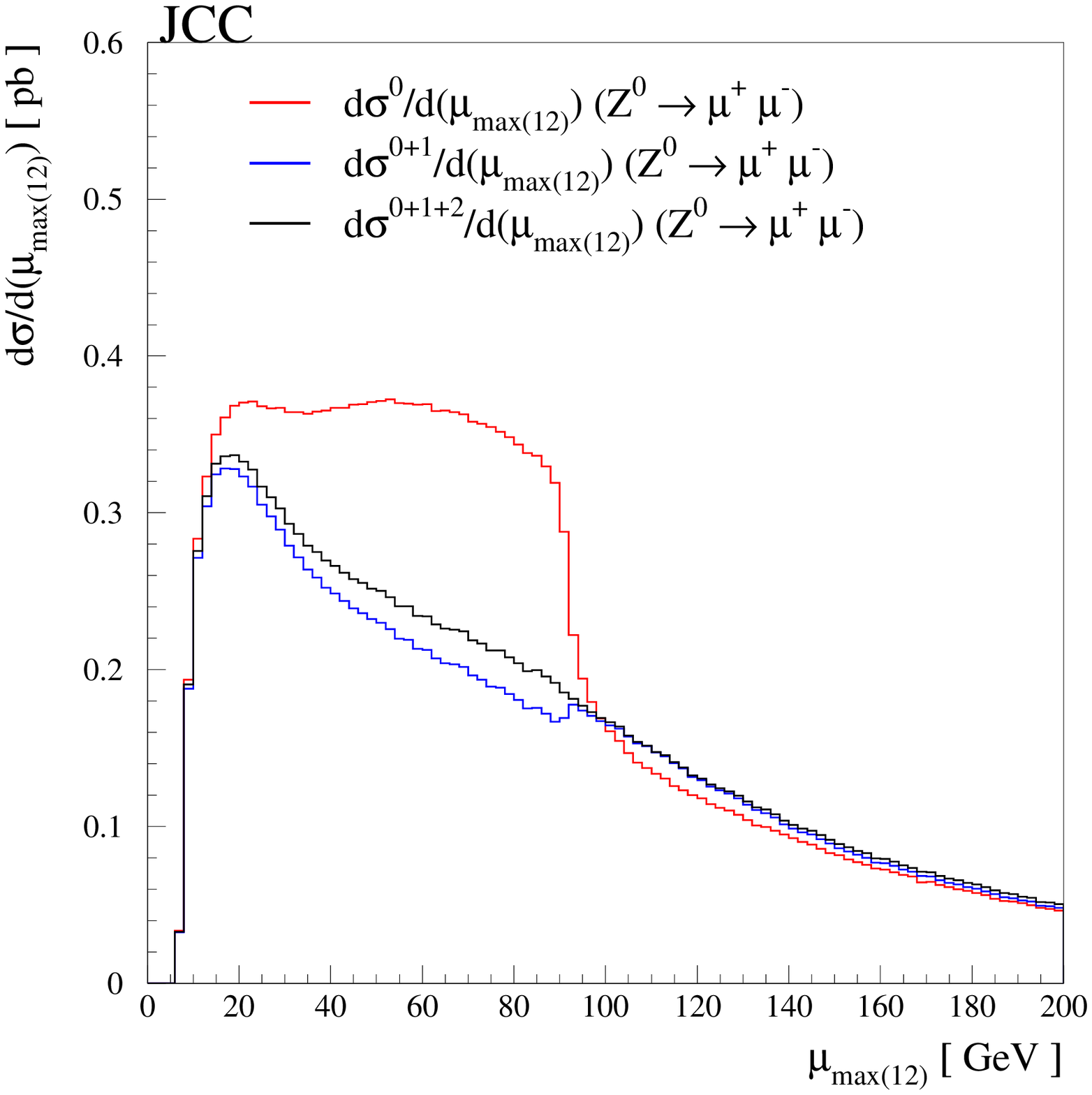,width=5.5cm}
\caption{ The full order $\alpha_s^2$ results 
using the CTEQ6L1 (top) and JCC PDF sets (bottom), showing the
distribution with respect to the virtuality/scale
related to first gluon and b-quark ($\mu_1$) (left) and maximal value
of $\mu_1$ $\mu_2$ (virtuality/scale of the second gluon and
anti-b quark) (right). As predicted the summed contribution with
overlap removal gives a smooth distribution despite the sharp cutoff at
$\mu = \mu_F = M_Z$. \label{f:alpha2}}
}

This is however not the case once we go to the full $\alpha_s^2$ order
where, as already stated, all the incoming b-quarks are showered back
to gluons. Here four possible scale choices exist $\mu_{i=1,4} =
-(p(g_{1,2})-p(b,\bar{b}))^2 + m_b^2$ and multiple overlap removal
terms can contribute. We thus chose to plot the virtualities/scales
related to first gluon and b-quark ($\mu_1$) and second gluon and
anti-b quark ($\mu_2$) and their combinations; all other permutations
in computing the virtualities give identical predictions. The
one-dimensional plots, as presented in Figure \ref{f:alpha2}, again
confirm the predictions of our approach resulting in a smooth
distribution from the combination of different order
contributions. The contributions to the total cross-sections are given
in Table \ref{t:xsect}. 
\TABLE{
\newcommand{\lstrut}{{$\strut\atop\strut$}}
  \caption { The process cross-sections for the leading order $\alpha_s^0$
  contribution integrated cross-section $\sigma^0$, order $\alpha_s^1$
  contribution $\sigma^1$ including the subtraction terms and order
$\alpha_s^2$ contribution $\sigma^2$, also with full overlap removal,
are shown for the $Z^0 \to \mu^+ \mu^- $ decay channel in the LHC
  environment (proton-proton collisions at $\rm \sqrt{s}$ = 14 TeV)
  are listed.  The b-quark mass is set to $m_b = 4.8$ GeV and the
  factorization and renormalization scales are set to the $Z^0$
  invariant mass squared. In addition, the perturbative (order $\alpha_s^{2}$)
  $gg \to Z b \bar{b} \to \mu^+ \mu^- b \bar{b}$ process cross-section
  is shown for comparison.  The cross-sections are given for the LO
  CTEQ6L1 \cite{Pumplin:2002vw} and the derived JCC PDFs. In the Monte-Carlo event generation
 procedure the  next-to-leading process weights are combined with the subtraction
  weights on the event-by-event basis as described in the text.
  \label{t:xsect}}

\begin{tabular}{lcc}
\hline
Process & $ \sigma_{\mathrm{CTEQ6L1,\mu_F = m_Z}}$ $\rm [ pb ]$ & $ \sigma_{\mathrm{JCC,\mu_F = m_Z}}$  $\rm [ pb ]$ \\
\hline
$\sigma^0$ & 64.4 & 44.8 \\
\hline
$\sigma^1$ & -10.7  & -8.9 \\
\hline
$\sigma^2$ & 2.0  & 2.0 \\
\hline
$\Sigma_i \sigma^i$ & 55.7 & 37.9 \\
\hline
$gg \to Z  b \bar{b} \to \mu^+ \mu^- b \bar{b}$ & 22.9 & 22.9 \\
\hline
\end{tabular}
}

One can observe that the contribution of the order
$\alpha_s^2$ is small both in the absolute normalization and its
contribution to the one-dimensional projections. Its impact and
importance is better observable plotting the two dimensional
($\mu_1,\mu_2$) differential cross-section plots presented in Figures
\ref{f:alpha2d},\ref{f:alpha2d1}. Apart from contributing to its `exclusive'
region of high $\mu_{1,2} >> \mu_F$ it contains a significant
correction in the region $\mu_1 \simeq \mu_2$ throughout the
$\mu_{1,2}$ value range. Upon reflection this is to be expected since
the perturbative-level calculation (and thus the full combined result)
should not be biased around the  $\mu_1 \simeq \mu_2$ region
whereas the parton-showering approach by definition \emph{is biased}
in this region, especially at values close to $\mu_F$. The probability
of having \emph{both} $\mu_{1,2}$ large is the square of probability
to obtain a single large $\mu$ value and its value is thus supposed to be low, with the
observable dip in the $\mu_{1,2}$ distributions Figures
\ref{f:alpha2d},\ref{f:alpha2d1} which the order $\alpha_s^2$ contribution
fills up as expected.
\FIGURE{
     \hspace{0.7cm}\epsfig{file=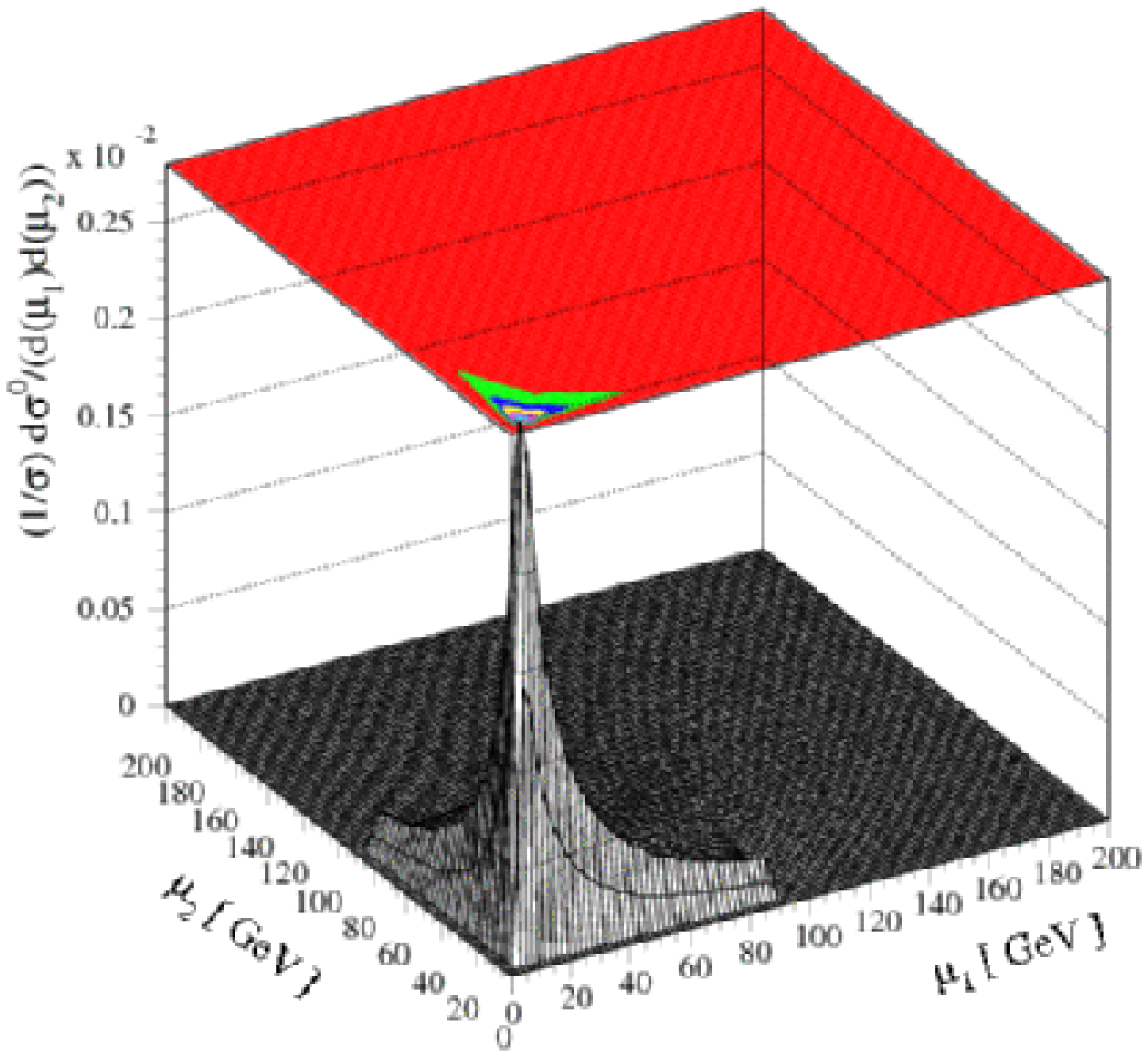,width=6.5cm}
     \epsfig{file=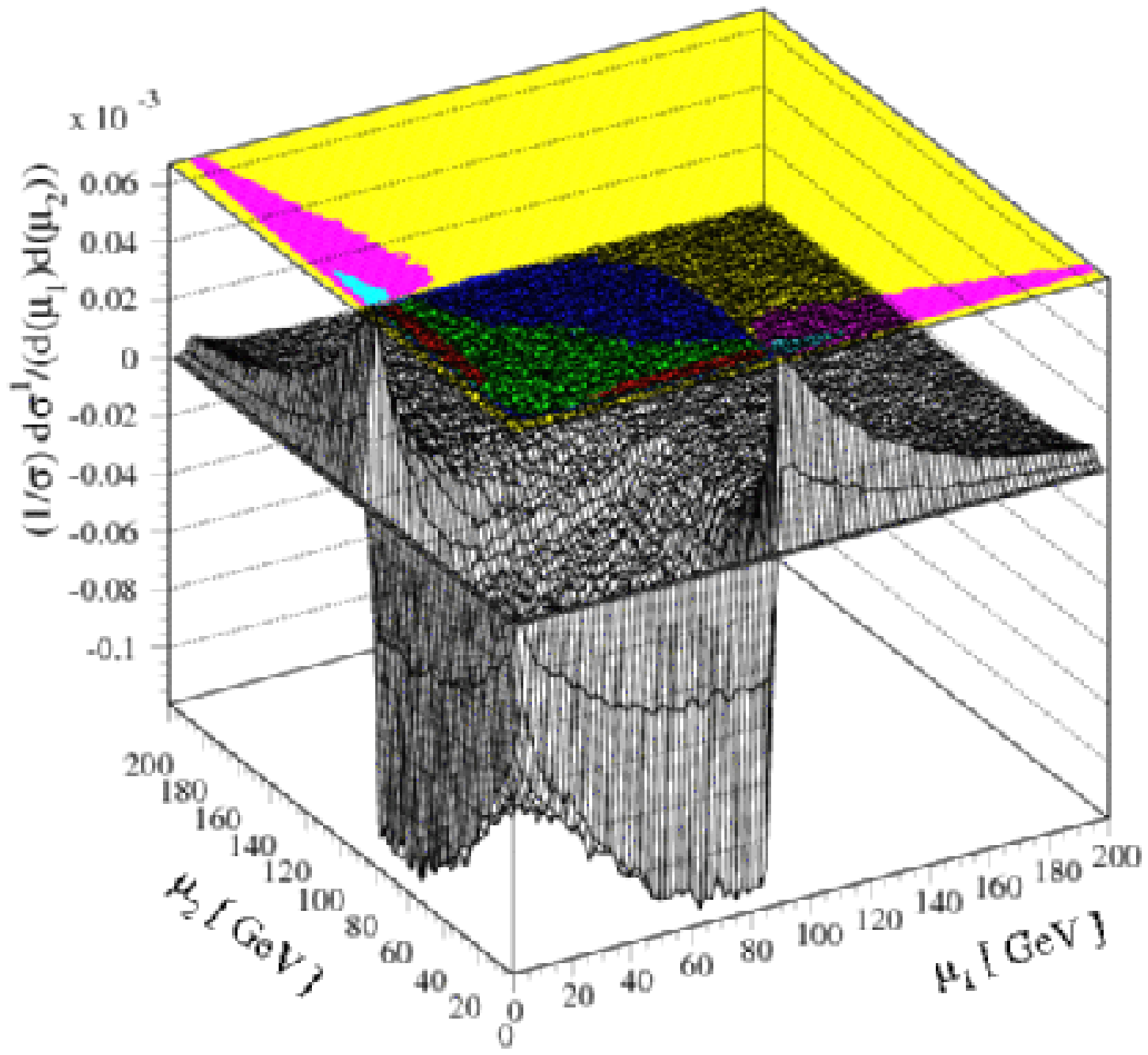,width=6.5cm}\\
     \epsfig{file=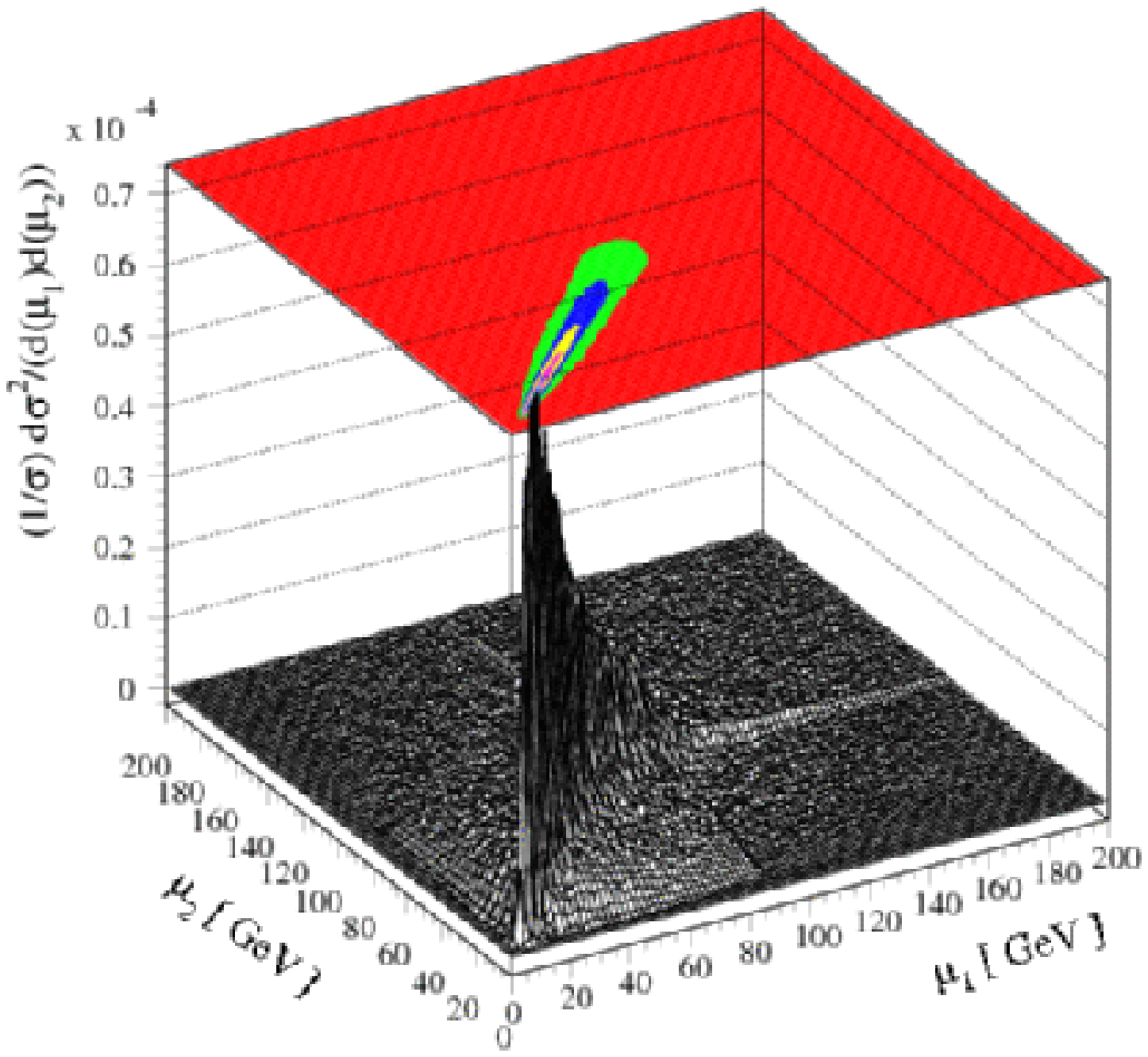,width=6.5cm}
     \epsfig{file=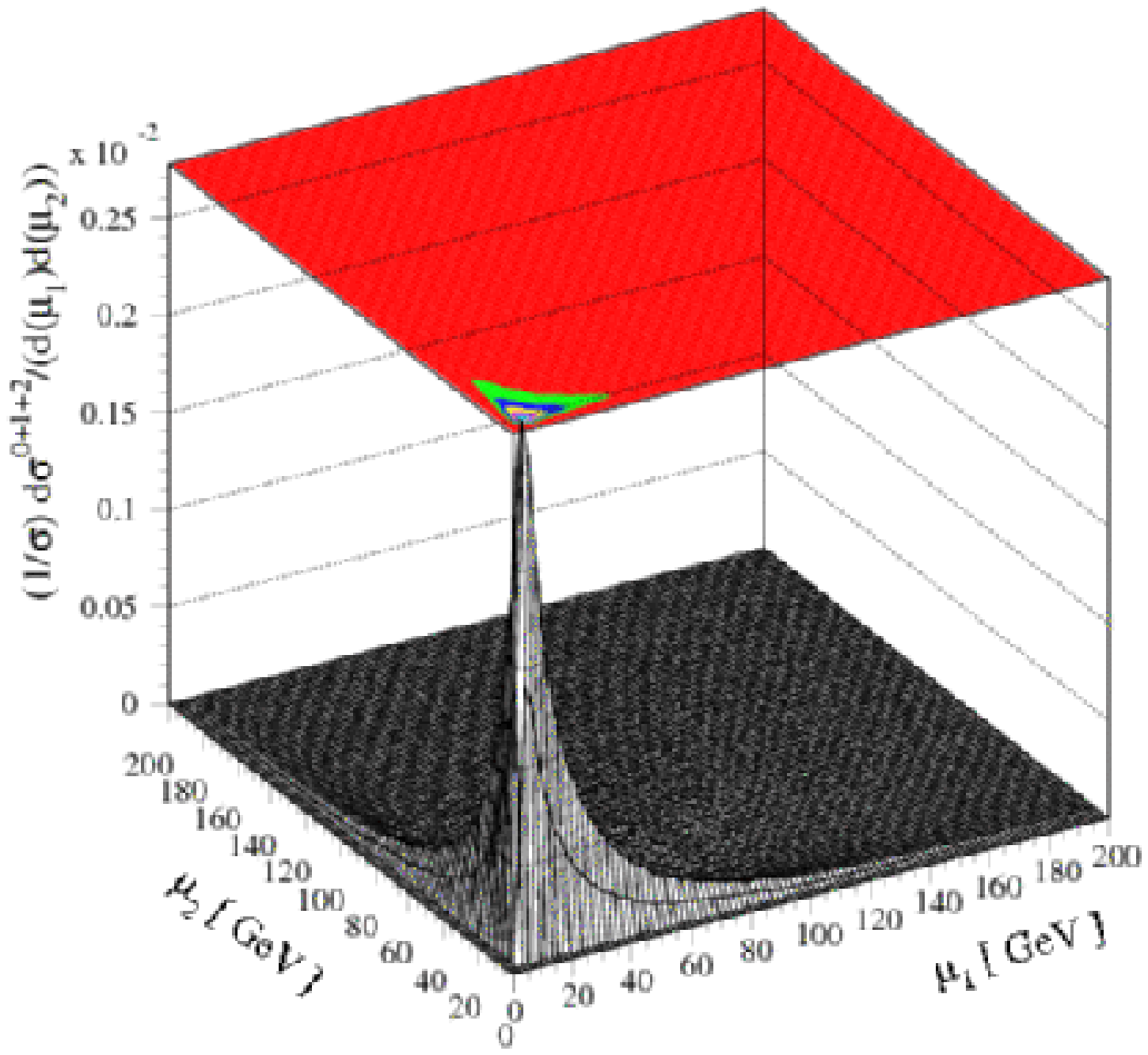,width=6.5cm}
\caption{ The full order $\alpha_s^2$ results 
using the JCC PDF sets, showing the
(normalized) distributions with respect to the virtuality/scale related
to first gluon and b-quark ($\mu_1$) and virtuality/scale of the
second gluon and anti-b quark) ($\mu_2$) for separate contributions
including their subtraction terms. On the bottom right the sum of all
contributions is shown to give a smooth distribution over the whole
region as expected. \label{f:alpha2d}} }

\FIGURE{
     \hspace{0.7cm}\epsfig{file=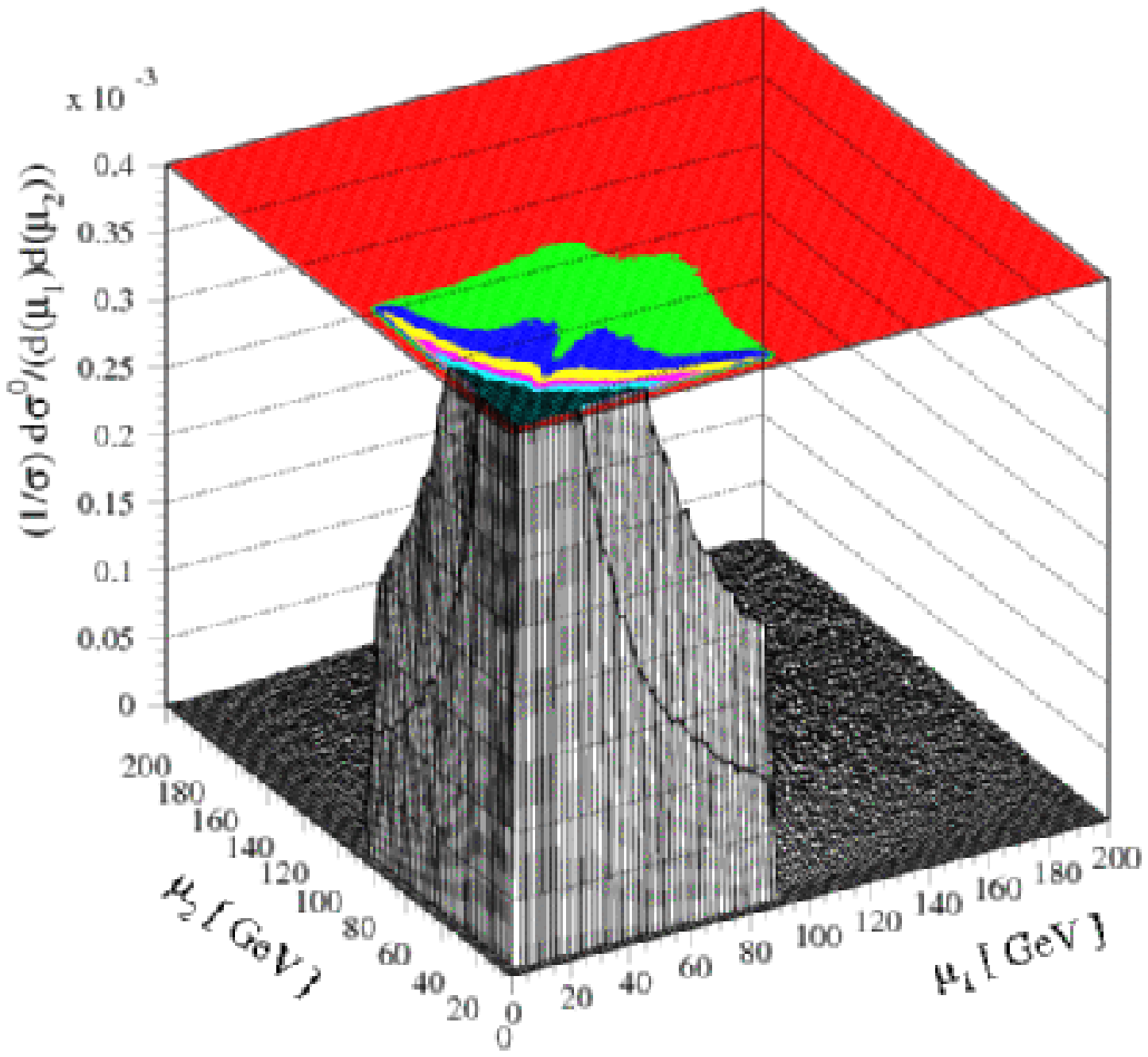,width=6.5cm}
     \epsfig{file=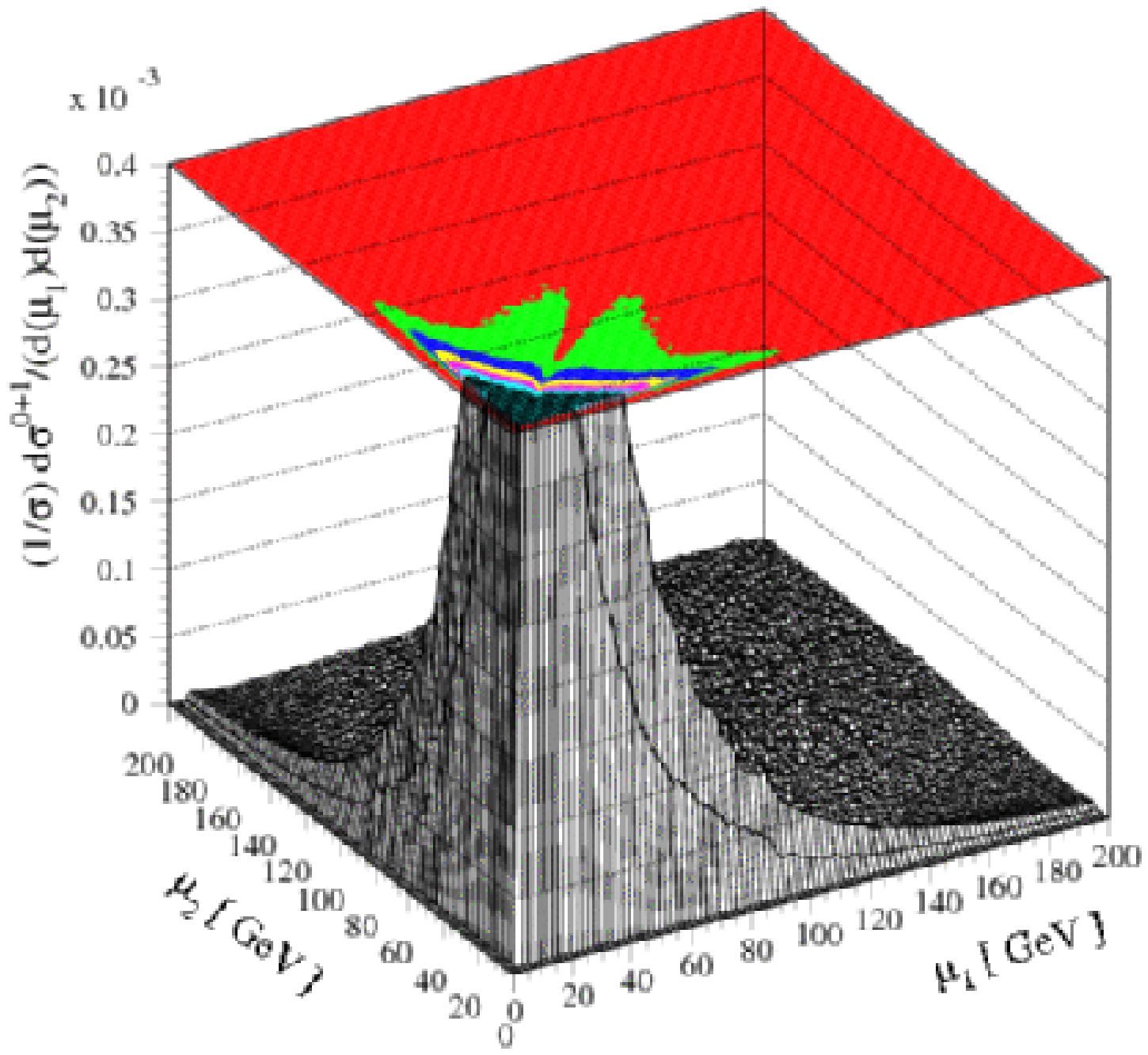,width=6.5cm}\\
     \epsfig{file=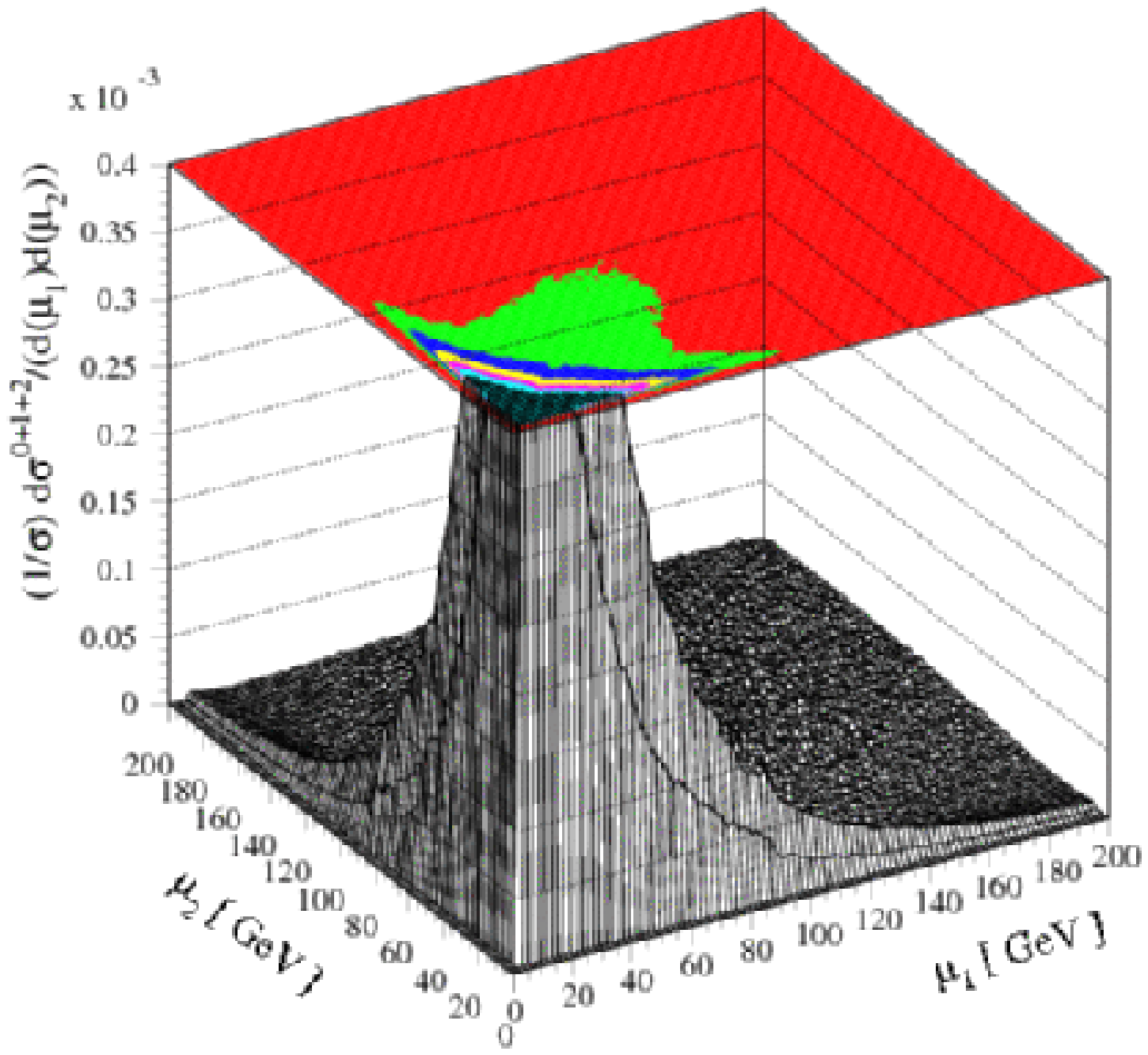,width=6.5cm}
\caption{ The full order $\alpha_s^2$ results 
using the JCC PDF sets, showing the
(normalized) distributions with respect to the virtuality/scale related
to first gluon and b-quark ($\mu_1$) and virtuality/scale of the
second gluon and anti-b quark) ($\mu_2$) for gradual combinations of contributions
of order $\alpha_s^{0,1,2}$ including their subtraction terms. Note
especially the disappearance of the dip in the region $\mu_1 \simeq
\mu_2$ as detailed in the text.\label{f:alpha2d1}} }

The kinematic quantity of interest is also the impact of subsequent
corrections on the transverse momentum of the $Z^0$ boson and b-quarks
as shown in the Figure \ref{f:kinematics}. Again, the
order $\alpha_s^2$ contribution seems to be comparatively small but
is of importance in the very high transverse momentum
regions. As it might be, one needs the full order $\alpha_s^2$ procedure to
achieve the exclusive final state containing two b-quarks and formally
obtain the full symmetry in the procedure.   
\FIGURE{
     \hspace{0.7cm}\epsfig{file=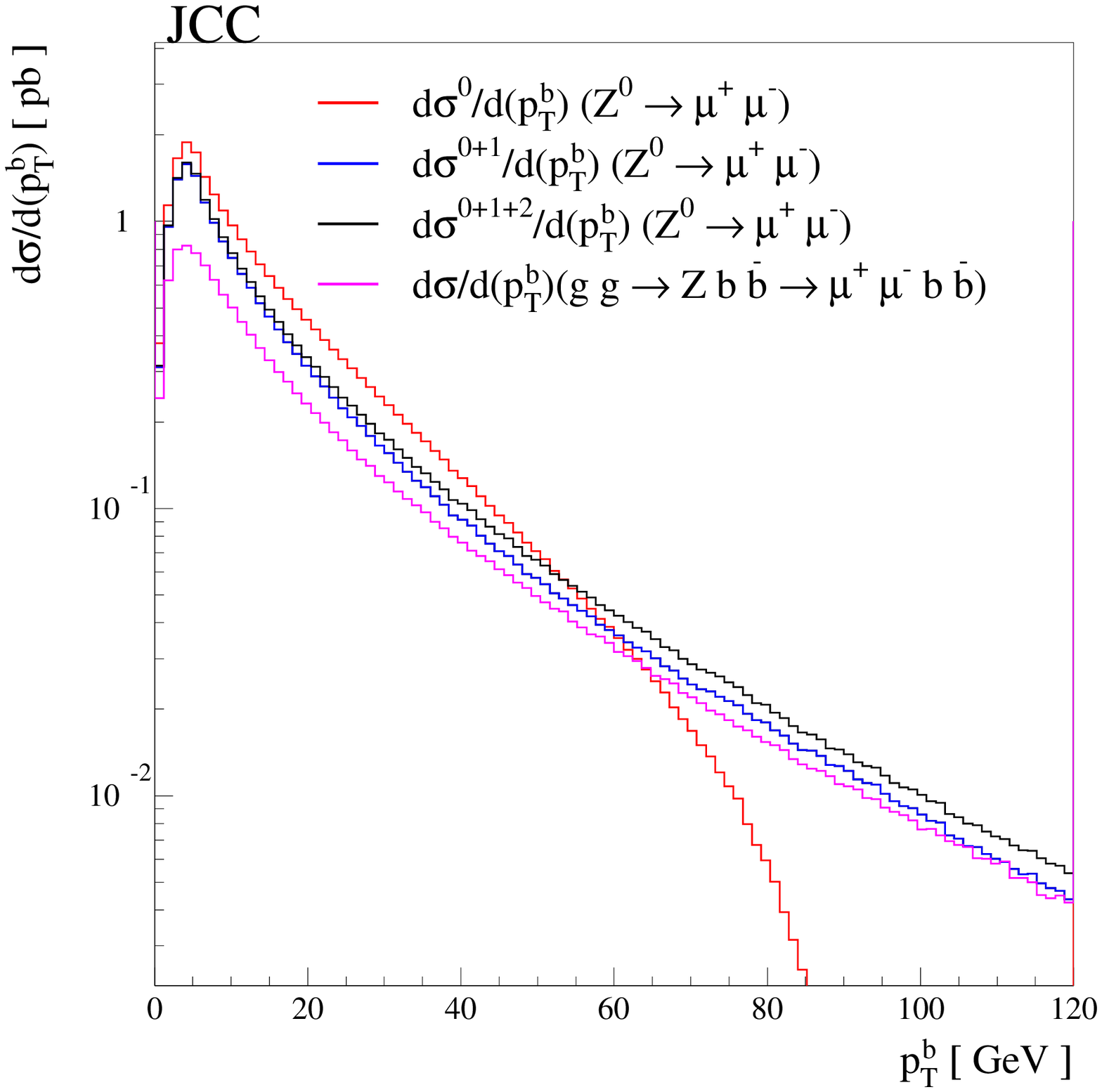,width=6.5cm}
     \epsfig{file=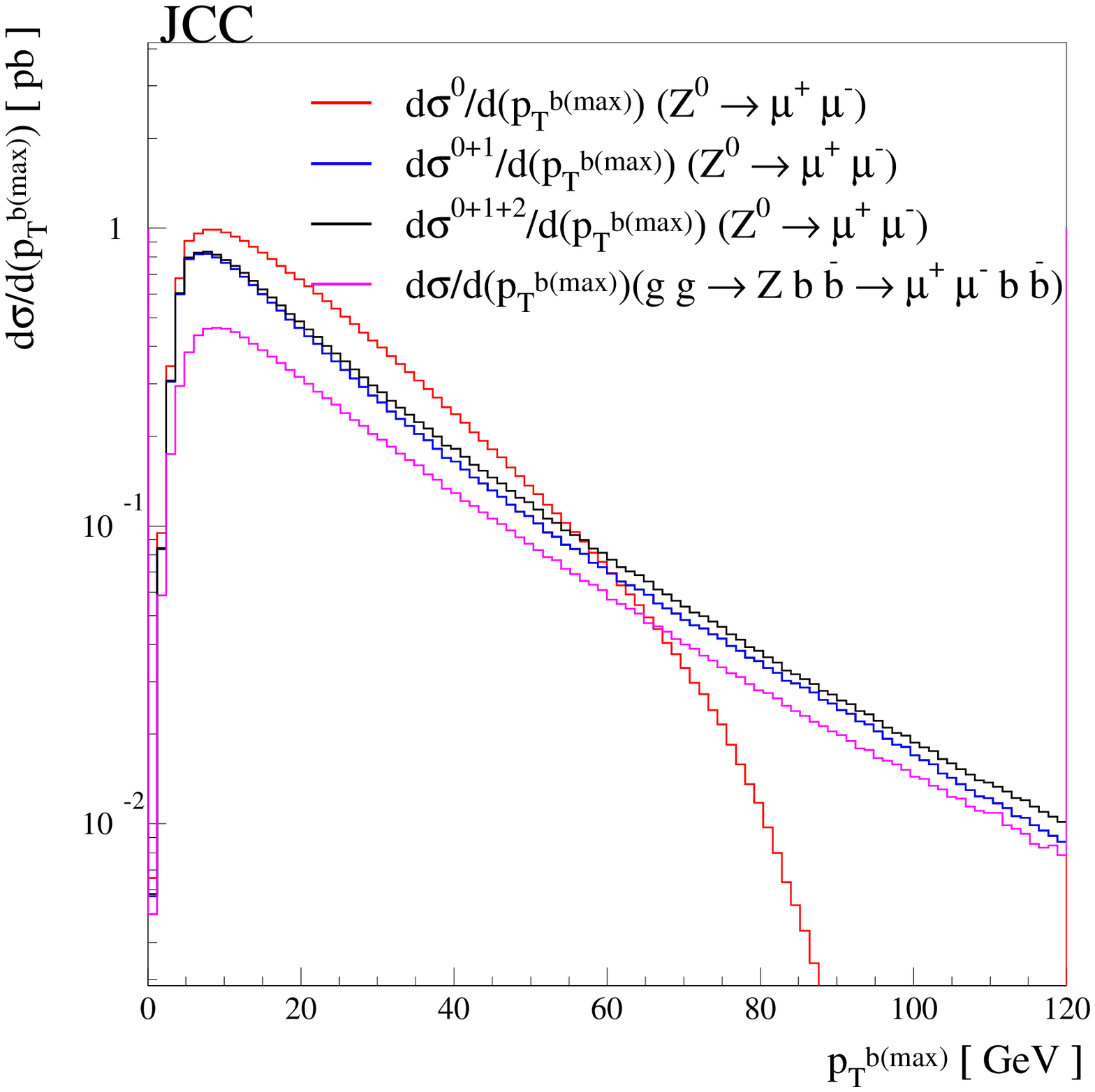,width=6.5cm}\\
     \epsfig{file=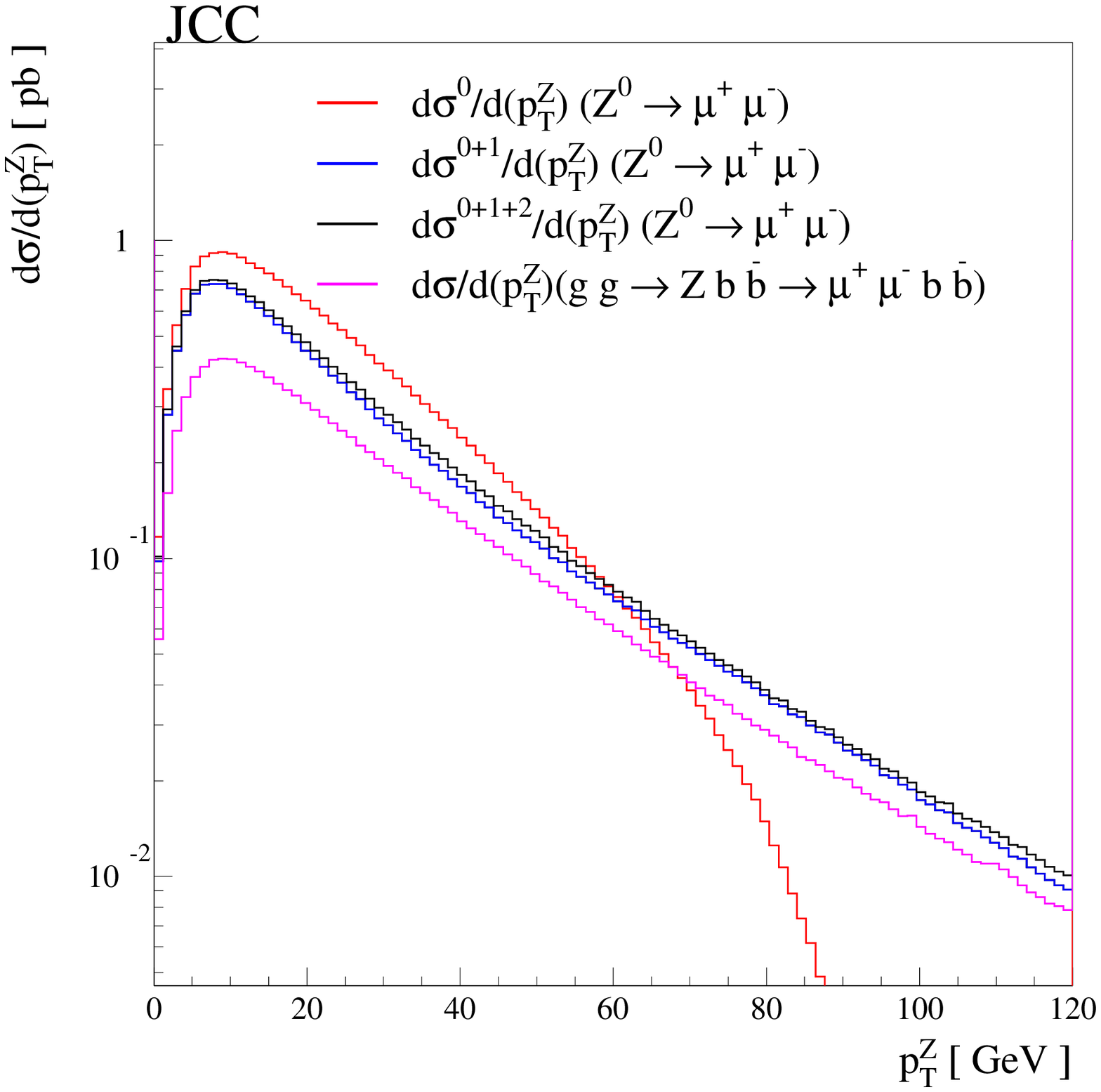,width=6.5cm}
\caption{ The full order $\alpha_s^2$ results 
using the JCC PDF sets, showing the differential cross-sections with
respect to the b-quark transverse momentum (left), maximal b-quark
transverse momentum (middle) and Z boson transverse momentum
(right). The perturbative calculation of order $\alpha_s^2$ without
subtractions is shown separately for comparison (magenta line).
\label{f:kinematics}} }

\section{Conclusions and Further Perspective}

It is instructive to compare our procedure to the procedure developed
in the MC@NLO \cite{Frixione:2002ik} framework. The MC@NLO Monte-Carlo
generator incorporates an
approach which results in double-counting removal both in the initial
and final state --- using massless quarks in the initial state --- at full
NLO, including virtual corrections, while our approach is currently
developed to use only tree-level matrix elements but can be used
iteratively to at least $\alpha_s^2$  as shown in the given
paper. Still, for our approach to be consistent, it should be
compatible to the MC@NLO formalism and it will be shown this is indeed
the case. MC@NLO is in its formalism following the paradigm of
'minimal intrusion' into the (showering) Monte-Carlo and is being
interfaced to HERWIG Monte-Carlo generator \cite{Corcella:2002jc} with the kinematic
translations adapted to its showering (evolution parameters
etc \ldots). Fortunately, it turns out one does not need to go into
details to show the consistency of the two approaches; it is enough to
use the `master modified subtraction equation' of the MC@NLO toy model in
\cite{Frixione:2002ik}, \ie Equation~(3.20), which states: 
\begin{eqnarray}
\left(\frac{d\sigma}{dO}\right)_{\rm msub} &=& \int\limits_0^1\, dx
\biggl[ I_{\rm MC}(0,x_M(x)) \frac{a[ R(x)-B Q(x)]}{x} \notag \\
&+& I_{\rm MC}(0,1)\left( B\, + \, aV\, + \frac{aB[Q(x) -1 ]}{x}\right) \biggr],
\label{f:mcatnlo}
\end{eqnarray}
where $B$ represents the LO Born term, $a V$ the finite part of the
virtual corrections, $a R(x)/x$ the unsubtracted real corrections and
$Q(x)$ is the Sudakov evolution kernel. The MC `term',
\begin{equation}
I_{\rm MC}(0,x_M) \frac{a B Q(x)}{x},
\end{equation}
equivalent to our subtraction terms, is in the above case \emph{both
subtracted and added} to the NLO cross-section expression to obtain
two separately finite contributions. As stated, in MC@NLO the shower
evolution is assumed fixed/pre-defined by the Monte Carlo program (defining the $x$ as
evolution variable and $Q(x)$ as the showering kernel) while the
subtraction scheme at NLO, which removes the divergences and defines
its own subtraction kernels is chosen separately (\eg $\rm \overline{MS}
$). For simplicity in the MC@NLO toy model the subtraction kernel is set
simply to one and the numerator $ a B [Q(x) -1 ]$ in last fraction in
Eq. \ref{f:mcatnlo} represents precisely the difference between the
shower evolution kernel (scheme) and the NLO subtraction kernel
(scheme).

\emph{If} one instead chooses another approach and \emph{adjusts} the
shower and NLO subtraction schemes to match, \ie
\begin{equation}
[Q(x) -1] \to [Q_{\rm shower} - Q_{\rm NLO}] \to 0,
\end{equation} one gets instead:
\begin{eqnarray}
\left(\frac{d\bar{\sigma}}{dO}\right)_{\rm msub} &=& \int\limits_0^1\, dx
\biggl[ I_{\rm MC}(0,x_M(x)) \frac{a[ R(x)-B Q(x)]}{x} \notag \\
&+& I_{\rm MC}(0,1)\left( B\, + \, aV\, \right) \biggr] ,
\label{f:mcatnlon}
\end{eqnarray}
where the MC term is present only in subtraction from the real
contribution but the NLO subtraction scheme is \emph{re-defined} to
match the showering scheme. If one disregards the finite virtual
correction (which affects only the normalization) and particle masses this matches exactly the
formula of our approach. It also becomes clear that the parton
distribution functions need to be re-defined since we are no longer
in the $\rm \overline{MS}$ scheme (or any other standard scheme) but in a
shower-governed scheme as clearly explained by Collins \emph{et al}
\cite{Collins:2000qd,Chen:2001nf}.  

The above comparison also shows the current limits of this approach:
Our method assumes that there are no other divergent terms in the
cross-sections apart from the mass divergences; furthermore, the
expressions for shower scheme $\rm JCC$ parton distribution functions
have so far been developed only for quarks. Consequently, if one wants
to incorporate the presence of \eg gluon radiation/splits in the hard
process this approach would need to be extended, combined with other
approaches (\eg the addition of light `jet-objects' through an
algorithm like Vincia \cite{Giele:2007di} or dipole showers
\cite{Schumann:2007mg}) or possibly even take new directions
\cite{Collins:2004vq}. Nevertheless, facing the stated limitations,
the presented procedure is thus shown to be consistent with MC@NLO
approach at NLO order ($\alpha_s^1$ corrections), while it does not
include virtual (normalization) corrections.

Another point of interest is to compare our approach to the one of
L-CKKW \cite{Catani:2001cc},\cite{Lonnblad:2001iq} as the most
advanced parton shower and matrix element matching algorithm so far. While
L-CKKW defines a very sophisticated interpolation scheme it only picks
the \emph{dominant} collinear topology on an event-by-event basis
by virtue of the employed jet clustering; in contrast, in our approach
more than one collinearity can be accounted for and subtracted for
each event. Furthermore, L-CKKW, while removing the double counting,
was not (yet) shown to be formally correct in case of hadrons as the
colliding particles \cite{Alwall:2007fs}.  On the other hand, the
 procedure described in this paper currently works only for initial state splits
involving heavy quarks; in multi-particle final states the heavy quark
lines are generally expected to be comparatively few. This would
lead to a very complex procedure if light quarks/gluons were added,
while it might not be necessary from the practical perspective since
L-CKKW seems to be describing the multi-light-jet final states very
well and could probably also be combined as a continuation of the procedure 
presented in this paper.

The near-future plans are to extend this approach to the heavy quark
production in the final state gluon splits $ g \to H \bar{H}$ and to
implement the same formalism in the associated $H b \bar{b}$
production which does in itself not need any further development of
the formalism. There are certainly also plans to compare our approach
with other \zbb implementations and its impact to the LHC predictions
is certainly envisaged but requires further work at the level of mock
experimental LHC analysis and is thus digressing from the scope of the
current paper.

\appendix

\section{Derivation of the Hard Cross Section Expressions and
  Subtraction Terms \label{s:app}}

The hard cross-section expressions and corresponding subtraction terms
can according to \cite{Olness:1997yc} be 
actually be derived from the Factorization Theorem itself by  using 
the Factorization Theorem \emph{at the parton level} and doing power
counting of $\alpha_s$. Specifically, at a given order in $\alpha_s^n$
the factorization Theorem implies that the perturbative pQCD  cross
section $\rm \sigma^{(n)}_{ab \to X}$ involving initial state partons $\rm a,b$ is 
related to its corresponding \emph{hard} cross section $\rm
\hat{\sigma}^{(n)}_{ab \to X}$ (containing no mass singularities and
being completely infra-red safe) 
by the modified Factorization Theorem\footnote{In the paper
  \cite{Olness:1997yc} also final state evolution (\ie fragmentation
  functions) are considered but in this paper we limit ourselves to
  the initial state evolution only, as stated several times in the
  paper. Subsequently, our choice of processes considered is constrained
  by this limitation.}:
\begin{equation}
\sigma^{(n)}_{ab \to X} = \sum_{c,d} f^{(n_1)}_{c/a} \otimes \hat{\sigma}^{(n_2)}_{cd \to X} \otimes
f^{(n_3)}_{d/b},
\label{e:rfac}
\end{equation}
where the sum of $n_i$ terms gives the correct order $n$, \ie
$n=n_1+n_2+n_3$.  The above equation differs from the regular Factorization Theorem
expression in two respects:
\begin{itemize}
\item The parton distributions are in this case relative to an
  on-shell \emph{parton} target.
\item These parton-level distributions are expanded in powers of
  $\alpha_s$, with $f^{(n_1)}_{c/a}$ denoting the term of order
  $\alpha_s^{n_1}$ in the distribution function expansion. 
\end{itemize}
Furthermore, we limit ourselves to the case where at least one of the
evolved incoming partons $c,d$ is massive with mass $M_H$. 
Using the ACOT scheme \cite{Aivazis:1993kh,Aivazis:1993pi}, the above parton densities are
calculated in the $\rm \overline{MS}$ scheme in the region $\mu > M_H$ at
the first two orders in $\alpha_s$ as:
\begin{eqnarray}
f_{j/i}^{(0)}(\xi) &=& \delta_i^j \delta(\xi - 1)\\
f_{H/i}^{(1)}(\xi,\mu) &=& \frac{\alpha_s(\mu)}{2\pi} P_{i \to H}(\xi)
\ln\left(\frac{\mu^2}{m_H^2}\right), \\
f_{g/g}^{(1)}(\xi,\mu) &=& \frac{\alpha_s(\mu)}{2\pi} \delta(\xi-1) \ln\left(\frac{\mu^2}{m_H^2}\right),
\end{eqnarray}
where $i,j=\{g,H \}$ and $P_{i \to H}(\xi)$ are the (usual) first order splitting kernels
for (heavy) quark production. In addition, in this scheme, the pQCD
cross-section $\sigma^{(n)}_{ab \to X}$ should already have been
regularized using the $\rm \overline{MS}$ scheme and the collinear
singularities should have already been subtracted resulting in a
finite cross section with only the
mass singularities (meaning, at finite $M_H$, large logarithms
$\ln(\mu^2/m_H^2)$) still present. In the calculation the light quark
masses should be set to zero and the $M_H$ kept at non-zero
value throughout the calculation (contrary to the conventional
approach where $M_H$ is eventually set to zero). The hard cross-section $\rm \hat{\sigma}^{(n)}_{ab \to X}$ is
then obtained by iteratively(recursively) inverting the Equation \ref{e:rfac}  (c.f.
\cite{Olness:1997yc}).

Let us illustrate how this approach was applied in this paper by
explicitly considering the \zbb process. At order $\alpha_s^2$ (n=2)
we get from equation \ref{e:rfac}:  
\begin{eqnarray}
\sigma^{(2)}_{\zbb } &=& f_{g/g}^{(0)}\, \otimes \, \hat{\sigma}^{(2)}_{\zbb }\,
\otimes \, f_{g/g}^{(0)} \label{e:rzbb2} \\
&+& \sum\limits_{H=b,\bar{b}} f^{(1)}_{H/g}(\mu_{1H})\, \otimes \, 
\hat{\sigma}^{(1)}_{\rm H g \to \zz H }\,\otimes \,f_{g/g}^{(0)} \notag\\
&+& \sum\limits_{H=b,\bar{b}}  f_{g/g}^{(0)}\, \otimes \, \hat{\sigma}^{(1)}_{\rm g \bar{H}
\to \zz \bar{H} }\,\otimes \,  f^{(1)}_{\bar{H}/g}(\mu_{2\bar{H}}) \notag \\
&+& \sum\limits_{H=b,\bar{b}} f^{(1)}_{H/g}(\mu_{1H})\, \otimes \,
\hat{\sigma}^{(0)}_{\rm H \bar{H} \to \zz }\, \otimes \,
f^{(1)}_{\bar{H}/g}(\mu_{2\bar{H}}),\, \notag
\end{eqnarray}
which exhausts the possible combinations. The order n=1,2 hard cross
sections $\hat{\sigma}^{(0,1)}$ still need to be determined so we move
our the procedure first one order down to n=1 and obtain ($H=b,\bar{b}$):
\begin{eqnarray}
\sigma^{(1)}_{\rm H g \to \zz H } &=& f_{H/g}^{(0)}\, \otimes \,
\hat{\sigma}^{(1)}_{\rm H g \to  \zz H }\, \otimes \,
f_{g/g}^{(0)} \label{e:rzbb1} \\
&+& f_{H/g}^{(0)}\, \otimes \, \hat{\sigma}^{(0)}_{\rm H \bar{H} \to \zz }\,
\otimes \, f^{(1)}_{\bar{H}/g}(\mu_{2\bar{H}}),\, \notag\\
\sigma^{(1)}_{\rm g \bar{H}\to \zz \bar{H} } &=& f_{g/g}^{(0)}\, \otimes \,
\hat{\sigma}^{(1)}_{\rm g \bar{H} \to \zz \bar{H} } \,
 \otimes \, f_{\bar{H}/g}^{(0)}\,\notag \\
 &+& f^{(1)}_{H/g}(\mu_{1H})\, \otimes \, \hat{\sigma}^{(0)}_{\rm H \bar{H}  \to \zz }\,
\otimes \, f_{\bar{H}/g}^{(0)},\, \notag
\end{eqnarray}
which still cannot be inverted so we move subsequently to n=0 and obtain:
\begin{equation}
\sigma^{(0)}_{\rm H \bar{H} \to \zz } = f_{H/g}^{(0)}\,\otimes \, 
\hat{\sigma}^{(0)}_{\rm H \bar{H} \to \zz }\, \otimes \,f_{\bar{H}/g}^{(0)} = 
\hat{\sigma}^{(0)}_{\rm H \bar{H} \to \zz }, \label{e:rzbb0}
\end{equation}
in which case the hard and perturbative cross sections are
equal, meaning:
\begin{equation}
\hat{\sigma}^{(0)}_{\rm H \bar{H} \to \zz } = \sigma^{(0)}_{\rm H \bar{H} \to \zz }.  \label{e:szbb0}
\end{equation}
Inserting this result back into Equation \ref{e:rzbb1} one
can now express the hard cross-section:
\begin{eqnarray}
\hat{\sigma}^{(1)}_{\rm H g \to \zz H } &=& 
\sigma^{(1)}_{\rm H g \to  \zz H }\, \label{e:szbb1}  \\
&-&  \sigma^{(0)}_{\rm H \bar{H} \to \zz }\, \otimes \,
f^{(1)}_{\bar{H}/g}(\mu_{2\bar{H}}),\, \notag\\
\hat{\sigma}^{(1)}_{\rm g \bar{H}\to \zz \bar{H} } &=& 
\sigma^{(1)}_{\rm g \bar{H} \to \zz \bar{H} } \, \notag \\
&-& f^{(1)}_{H/g}(\mu_{1H})\, \otimes \, \sigma^{(0)}_{\rm H \bar{H} \to \zz },\, \notag
\end{eqnarray}
and now these results finally into Eq. \ref{e:rzbb2}, giving:
\begin{eqnarray}
\hat{\sigma}^{(2)}_{\zbb } &=& \sigma^{(2)}_{\zbb }\, \label{e:szbb2r} \\
&-& \sum\limits_{H=b,\bar{b}} f^{(1)}_{H/g}(\mu_{1H})\, \otimes \,
\left[ \sigma^{(1)}_{\rm H g \to  \zz H }\, -  \sigma^{(0)}_{\rm H
    \bar{H} \to \zz }\, \otimes \, f^{(1)}_{\bar{H}/g}(\mu_{2\bar{H}}) \right]  \, \notag \\
&-& \sum\limits_{H=b,\bar{b}} 
\left[ \sigma^{(1)}_{\rm g \bar{H} \to \zz \bar{H} } \, -
  f^{(1)}_{H/g}(\mu_{1H})\, \otimes \, \sigma^{(0)}_{\rm \bar{H} H \to \zz
  },\,\right] \, \otimes \, f^{(1)}_{\bar{H}/g}(\mu_{2\bar{H}}) \notag\\
&-& \sum\limits_{H=b,\bar{b}} f^{(1)}_{H/g}(\mu_{1H})\, \otimes \,
\sigma^{(0)}_{\rm H \bar{H} \to \zz }\, \otimes \,
f^{(1)}_{\bar{H}/g}(\mu_{2\bar{H}}),\, \notag \\
&=& \sigma^{(2)}_{\zbb }\, \label{e:szbb2} \\
&-& \sum\limits_{H=b,\bar{b}} f^{(1)}_{H/g}(\mu_{1H})\, \otimes \, \sigma^{(1)}_{\rm H g \to  \zz H } \, \notag \\
&-& \sum\limits_{H=b,\bar{b}}  \sigma^{(1)}_{\rm g \bar{H} \to \zz \bar{H} } \, \otimes \, f^{(1)}_{\bar{H}/g}(\mu_{2\bar{H}}) \notag\\
&+& \sum\limits_{H=b,\bar{b}} f^{(1)}_{H/g}(\mu_{1H})\, \otimes \,
\sigma^{(0)}_{\rm H \bar{H} \to \zz }\, \otimes \,
f^{(1)}_{\bar{H}/g}(\mu_{2\bar{H}}).\, \notag
\end{eqnarray}
As one can observe from Equations \ref{e:szbb0},\ref{e:szbb1} and \ref{e:szbb2} the
hard cross section can be expressed as the difference of the
perturbative cross section and a subtraction term:
\begin{equation}
\hat{\sigma}^{(n)} = \sigma^{(n)} - \sigma_{\rm subt}^{(n)}.
\end{equation}
Let us stress again that in this approach the $M_H$ is left at
non-zero value in all the above terms, contrary to the conventional
approach where $M_H$ is eventually set to zero.

It should be emphasized that in this paper only Born-level
(tree-level) pQCD calculations were used, thus implicitly assuming
that there are no divergences beyond massive ones present for the
process at hand. As a consequence, all contributions involving the
additional participation of (soft) gluons in the final state $X$, which would require
explicit NLO (loop) calculations, are not considered.

\end{document}